\documentclass[a4paper,fleqn]{cas-dc}

\usepackage[numbers]{natbib}
\usepackage{algorithm,algorithmic,subfigure,stfloats,float,threeparttable}

\def\tsc#1{\csdef{#1}{\textsc{\lowercase{#1}}\xspace}}
\tsc{WGM}
\tsc{QE}
\tsc{EP}
\tsc{PMS}
\tsc{BEC}
\tsc{DE}

\begin{document}
\let\WriteBookmarks\relax
\def\floatpagepagefraction{1}
\def\textpagefraction{.001}
\shorttitle{Building an intrusion detection system}
\shortauthors{Yuyang Zhou et~al.}

\title [mode = title]{Building an Efficient Intrusion Detection System Based on Feature Selection and Ensemble Classifier}

\author[1,2,3]{Yuyang Zhou}[type=editor,
                        auid=000,bioid=1,
                        orcid=0000-0001-8626-0468]
\ead{yyzhou@njnet.edu.cn}

\address[1]{School of Cyber Science and Engineering, Southeast University, Nanjing, China}
\address[2]{Key Laboratory of Computer Network and Information Integration, Ministry of Education, Nanjing, China}
\address[3]{Jiangsu Provincial Key Laboratory of Computer Network Technology, Southeast University, Nanjing, China}
\address[4]{National Key Laboratory of Science and Technology on Information System Security, Beijing, China}

\author[1,2,3]{Guang Cheng}[style=chinese]
\cormark[1]
\ead{gcheng@njnet.edu.cn}
\cortext[cor1]{Corresponding author}

\author[1,4]{Shanqing Jiang}[style=chinese]
\ead{sqjiang@njnet.edu.cn}

\author[1,2,3]{Mian Dai}[style=chinese]
\ead{mdai@njnet.edu.cn}

\begin{abstract}
Intrusion detection system (IDS) is one of extensively used techniques in a network topology to safeguard the integrity and availability of sensitive assets in the protected systems. Although many supervised and unsupervised learning approaches from the field of machine learning have been used to increase the efficacy of IDSs, it is still a problem for existing intrusion detection algorithms to achieve good performance. First, lots of redundant and irrelevant data in high-dimensional datasets interfere with the classification process of an IDS. Second, an individual classifier may not perform well in the detection of each type of attacks. Third, many models are built for stale datasets, making them less adaptable for novel attacks. Thus, we propose a new intrusion detection framework in this paper, and this framework is based on the feature selection and ensemble learning techniques. In the first step, a heuristic algorithm called CFS-BA is proposed for dimensionality reduction, which selects the optimal subset based on the correlation between features. Then, we introduce an ensemble approach that combines C4.5, Random Forest (RF), and Forest by Penalizing Attributes (Forest PA) algorithms. Finally, voting technique is used to combine the probability distributions of the base learners for attack recognition. The experimental results, using NSL-KDD, AWID, and CIC-IDS2017 datasets, reveal that the proposed CFS-BA-Ensemble method is able to exhibit better performance than other related and state of the art approaches under several metrics.
\end{abstract}


\maketitle

\section{Introduction}

Nowadays, the applications of the Internet help society in many areas such as electronic communication, teaching, commerce, and entertainment, it has become a part of daily life of the people. However, cyber security has become vulnerable due to the massive expansion of the computer networks and rapid emergence of the intrusion incidents. The necessity of developing cyber security has attracted considerable attention from industry and academia around the world. Despite the use of different security applications, such as firewalls, malware prevention, data encryption, and user authentication, many organizations and enterprises fall victims to contemporary cyber-attacks~\cite{al2015data}. In order to sneak into the system, attackers might deliberately exploit the vulnerabilities of the target system and launch different types of attacks, which may lead to the leakage of private information.

As technology is rolling out, these attacks threaten the confidentiality, integrity, and availability of cyber systems all the time. Therefore, it is necessary to introduce intrusion detection systems (IDSs)~\cite{elhag2015combination,wang2017strategic,wang2016attack,wang2016game} to protect systems from a variety of attacks. To be more specific, IDSs are widely deployed in various distributed systems, perceiving the malicious intrusions and then taking rapid countermeasures to prevent further infections and spread. In general, IDSs can be classified into two major categories based on detection mechanisms: anomaly and misuse detection~\cite{joldzic2016transparent}. In detail, anomaly detection is designed to detect malicious actions through identifying deviations from a normal profile behavior. Such IDSs perform better at detecting novel types of attacks, however, they could not avoid a high false positive (FP) rate~\cite{papamartzivanos2018dendron}. On the other hand, based on known patterns, misuse detection can effectively distinguish legitimate instances from the malicious ones~\cite{kim2016long}. Although this kind of IDSs is reliable for detecting known attacks, it cannot identify unknown attacks or variations of known ones.

Unfortunately, as the attackers become more sophisticated, new threats and vulnerabilities emerge rapidly. On the one hand, the risk for critical infrastructures to be compromised significantly increases in short order. On the other hand, in order to detect and deal with novel attacks, a higher requirement for IDS has also been brought forward. Hence, many approaches have been researched and developed to improve the detection rate and performance of IDSs. One of them is Machine learning (ML)~\cite{du2018big,du2018differential,mishra2018detailed}, which can be applied for both anomaly and misuse detection models. By analyzing network traffic passing through central network nodes, an IDS not only needs to distinguish between benign and malicious traffic, but also infers the specific class of an attack occurring in the protected system.

However, in most instances, only a fraction of the traffic may indicate malicious behaviors while a network is flooded with normal traffic flows, which leads to the difficulty of identifying attacks with high Attack Detection Rate (ADR) while keeping the False Alarm Rate (FAR) low. There was one problem with the initial idea of applying ML in IDS, that is, a single classifier may not be strong enough to build a good IDS. Thus, researchers have come up with the idea of constructing ensemble classifiers for IDSs~\cite{feng2018dynamic,salo2019dimensionality}. In general, the main goal of ensemble learning is to combine a set of individual classifiers and then make a better classification decision about the object submitted at the input~\cite{pham2018improving}. For instance, training a single classifier on different subsets of an IDS dataset could produce different classification performances, however, an ensemble would average the output of multiple classifiers and therefore become a better option.

Moreover, the numerous attack types and network traffic attributes pose another challenge for ML as they expand the search space of the problem and lead to high computational and time complexity~\cite{aljawarneh2018anomaly}. Notably, feature selection has been proven to be a good solution for an IDS, which detects highly relevant features and eliminates useless ones with a minimum degradation of performance~\cite{hota2014decision,khammassi2017ga}. There are three main models that deal with feature selection: wrapper, filter, and embedded approaches. Information gain ratio based feature selection is one of classical filter algorithms, where information gain ratio represents a ratio of information gain to the intrinsic information. Although it solves the drawback of information gain and reduces a bias towards multi-valued attributes, however, it may be biased towards features with fewer values in some instances. Different from information gain ratio, correlation-based feature selection maximizes the relevance between the input features and the output and minimizes the redundancy of the selected features. This algorithm selects one feature at a time according to its strong correlation with outputs, which can be used to perform both attribute selection and tuple reduction flexibly.

In this paper, we propose a novel intrusion detection system to detect various types of attacks with high accuracy and efficiency. First, as a regular means of dimensionality reduction and redundancy elimination, a nature-inspired feature selection algorithm is proposed to retrieve a subset of the original features. Second, the imbalance between normal and malicious traffic has a negative effect on the accuracy of attack detection. To overcome this problem, our solution then utilizes ensemble classifier to reduce the bias among different training datasets. In this way, feature selection and ensemble classifier are combined to improve the stability and accuracy of the IDS with low computational and time complexity. Finally, an unbiased model can be generated to detect both popular and rare intrusive events. The major contributions of our work are summarized as follows:

\begin{itemize}
  \item We propose a novel methodology that combines the benefits of feature selection and ensemble classifier with the aim of providing efficient and accurate intrusion detection.
  \item In the context of feature selection, we provide CFS-BA based approach, which is used to assess the correlation of the selected features and beneficial for optimizing the efficiency of the training and testing phase.
  \item To increase the multi-class classification performance on unbalanced datasets, we introduce an ensemble approach by combining decisions from multiple classifiers (C4.5, RF, and Forest PA) into one by utilizing a vote classifier based on the average of probabilities (AOP) combination rule.
  \item The proposal is compared with exsiting methods on an extensive testbed comprising of three datasets, namely: NSL-KDD, AWID, and CIC-IDS2017. Experimental results show that the proposed solution surpasses equivalent methods in terms of Accuracy (Acc), F-Measure, and ADR classification metrics, while keeping FAR at acceptable levels.
\end{itemize}

The rest of the paper is organized as follows. In Section~\ref{section2}, we review the background information concerning IDSs. Then, the proposed methodology is given in Section~\ref{section3}, while in Section~\ref{section4} we provide the evaluation results through experiments and comparative analysis. Finally, the conclusion is presented in Section~\ref{section5}.

\section{Related work}\label{section2}
As a significant tool in computer based systems for ensuring cyber security, IDS constantly attracts the research community's attention. Although plenty of solutions have been proposed to improve the performance of IDS, in the context of this section, we only consider related work that falls under the ML based IDS, utilizes feature selection or ensemble classifier, and especially focuses on hybrid approaches.

\subsection{On feature selection techniques}
For purpose of reducing computational complexity, the technique of feature selection~\cite{maza2018feature,mi2018software}, that can be used as a pre-processing step in ML algorithms, aims to eliminate irrelevant features while preserving or even enhancing the performance of the IDS. In order to obtain more robust and effective classifier, Hota and Shrivas~\cite{hota2014decision} proposed a model that used different feature selection techniques to remove irrelevant features. The results indicate that C4.5 with information gain can achieve the highest accuracy with only 17 features for the NSL-KDD dataset. In addition, Khammassi and Krichen~\cite{khammassi2017ga} have applied  as a search strategy and logistic regression as a learning algorithm for network IDSs to choose the best subset. The results demonstrate that their method provides high detection rate with only 18 features for the KDDCup'99 and 20 features for the UNSW-NB15 dataset. Abdullah et al.~\cite{abdullah2018enhanced} also proposed a framework of IDS with selection of features within the NSL-KDD dataset that are based on dividing the input dataset into different subsets, and combining them using Information Gain (IG) filter.

\subsection{On ensemble classifiers}
Moreover, ensemble methods are machine learning techniques that combine several base models in order to reduce false positive rates and produce more accurate solutions than a single model would. Gaikwad and Thool~\cite{gaikwad2015intrusion} proposed a bagging ensemble method using REPTree as its base classifier, which takes less time to build the model and provides highest classification accuracy with lowest false positives on the NSL-KDD dataset. Jabbar et al.~\cite{jabbar2017cluster} proposed a cluster-based ensemble classifier for IDS, which is built with Alternating Decision Tree (ADTree) and k-Nearest Neighbor algorithm (kNN). The experimental results show that the proposed ensemble classifier outperforms other existing techniques in terms of accuracy and detection rate. In order to create a stronger learner, Paulauskas and Auskalnis~\cite{paulauskas2017analysis} proposed an ensemble model of four different base classifiers: J48, C5.0, Naive Bayes, and Partial Decision List (PART), which depends on the idea of combining multiple weaker learners. Results prove that their ensemble model produces more accurate results for an IDS. In order to mitigate malicious events, in particular botnet attacks in Internet of Things (IoT) networks, Moustafa et al.~\cite{moustafa2018ensemble} proposed new statistical flow features and developed an AdaBoost ensemble learning method to detect attacks effectively.

\subsection{On hybrid approaches}

Recently, many hybrid approaches using both feature selection and ensemble method have been produced to improve the performance of IDSs. Malik et al.~\cite{malik2015network} proposed a combination approach of Particle Swarm Optimization (PSO) and Random Forest (RF). More appropriate features for each class help the proposed model produce a higher accuracy along with low false positive rate compared with other algorithms. Pham et al.~\cite{pham2018improving} built a hybrid model, which utilizes gain ratio technique as feature selection and bagging to combine tree-based base classifiers. Experimental results show that the best performance was produced by the bagging model that used J48 as the base classifier and worked on 35-feature subset of the NSL-KDD dataset. Abdullah et al.~\cite{abdullah2018enhanced} also built an IDS using IG based feature selection and ensemble learning algorithms. The experiment on NSL-KDD dataset indicates that the highest accuracy obtained when using RF and PART as base classifiers under the product probability rule. In addition, Salo et al.~\cite{salo2019dimensionality} proposed a hybrid IDS which combines the feature selection approaches of IG and Principal Component Analysis (PCA) with an ensemble classifier based on Support Vector Machine (SVM), Instance-Based learning algorithms (IBK), and Multi-Layer Perceptron (MLP). A comparative analysis performed on several IDS datasets has proven that IG-PCA-Ensemble method exhibits better performance than the majority of existing approaches. Due to large-scale data produced from a massive network infrastructure, Khan et al.~\cite{khan2019scalable} proposed a scalable and hybrid IDS, which is based on Spark ML and Convolutional-LSTM (Conv-LSTM) network to employ the anomaly and misuse detection separately. Zhong et al.~\cite{zhong2020helad} also proposed a new anomaly detection model called HELAD, which is based on the Damped Incremental Statistics algorithm for feature selection and organic integration of multiple deep learning techniques for classification. In~\cite{tama2019tse}, a novel IDS based on hybrid feature selection and two-level classifier ensembles has been proposed, and experimental results show that it produces a significant improvement of the detection rate on the NSL-KDD and UNSW-NB15 datasets.

\section{Proposed methodology}\label{section3}
In order to increase the detection ability of IDS and prevent the service providers from attack, we propose an efficient ML-based IDS using a metaheuristic optimization algorithm based feature selection approach, and a vote classifier which is an ensemble of classifiers method. The AOP combination rule is integrated into the model for the decision step. During the experiments, 10-fold cross-validation (CV) approach is used to validate the performance of the model and classify benign traffic and various types of attacks.

\begin{figure*}
\centering
    \includegraphics[width=6.5in]{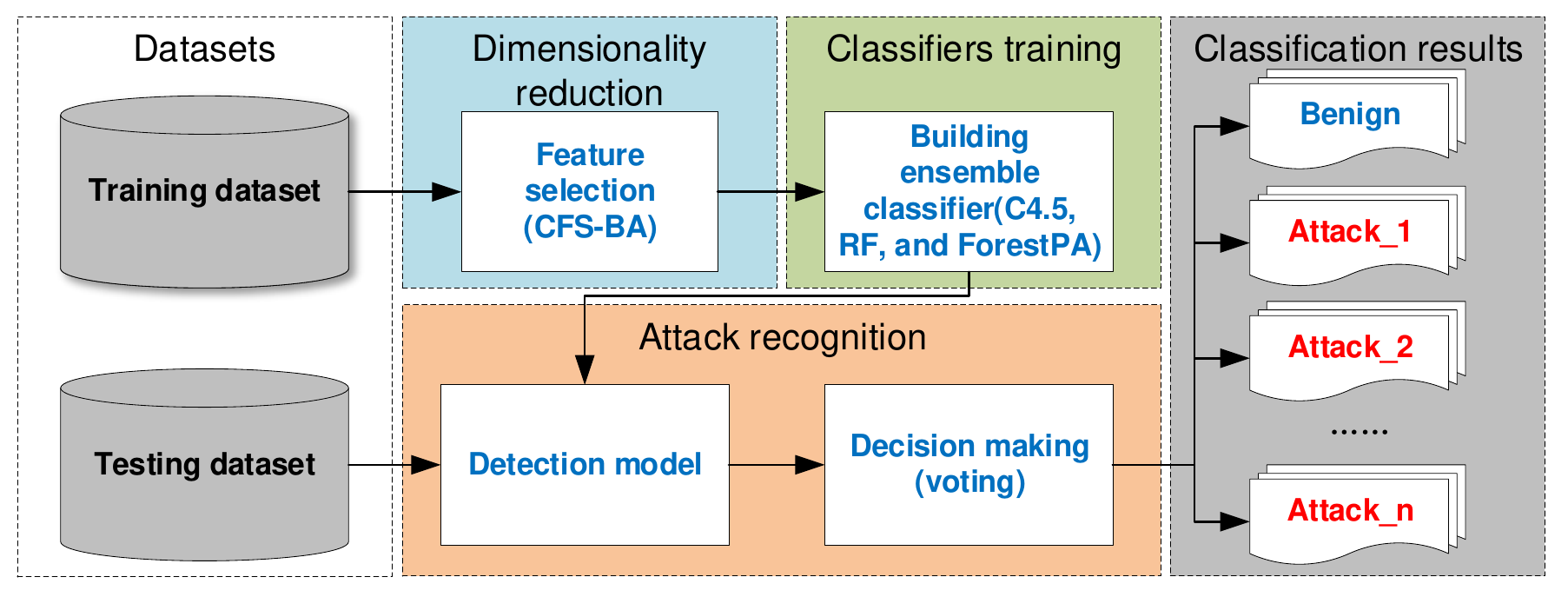}
    \caption{The framework of the proposed Feature selection-Ensemble model.}
    \label{fig:framework}
\end{figure*}

Fig.~\ref{fig:framework} demonstrates the detection framework of the proposed ML-based IDS, which consists of the following four main phases:

\begin{itemize}
  \item Datasets preprocessing: The first phase is to transform raw data into a format suitable for analysis by applying preprocessing to the original datasets.
  \item Dimensionality reduction: In order to overcome the problem of high-dimensional datasets, the feature selection approach based on CFS-BA is used to reduce the dimensionality of the datesets and select the most relevant features for each type of attacks.
  \item Classifiers training: For purpose of improving the accuracy of the IDS, we train three individual classifiers as base learners using C4.5, RF, and Forest PA, and build an ensemble classifier based on them.
  \item Attack recognition: The detection model is tested using a 10-fold cross-validation approach, and voting technique is used to combine the probability distributions of the base learners with the AOP combination rule to make classification decisions.
\end{itemize}

Finally, according to the results of the ensemble classifier, benign traffic and various intrusive events can be detected and classified with high classification accuracy. Detailed information about the framework is provided in Sections~\ref{subsection1}$-$\ref{subsection2}.

\subsection{Feature selection}\label{subsection1}
The aim of feature selection is to find a subset of the attributes from the original set which are representative enough for the data, and the attributions in the subset are highly relevant to the prediction. Feature selection approaches can be mainly categorized into wrapper, filter, and embedded approaches~\cite{hajisalem2018hybrid}. While filter approaches assess the relevance of the features from the dataset and the selection of the features is based on the statistics, the classification performance is used in wrapper approaches as a part of the feature subsets evaluation and selection processes. In contrast to wrapper approaches, embedded approaches are computationally less intensive because they incorporate an interaction between feature selection and learning process. Although embedded approaches integrate a regularised risk function to optimize the features designating parameters and the predictor parameters~\cite{bolon2016feature}, it is not easy to make a modification in the classification model to get higher performance~\cite{liu2005toward}.

Modern intrusion detection datasets inevitably contain plenty of redundant and irrelevant attributes~\cite{acharya2018iwd}, which lower the efficacy of data mining algorithms and cause uninterpretable results~\cite{chen2013classification}. Therefore, the first step in this study is to reduce the dimensionality and select the feature subset of the utilized dataset~\cite{salo2019dimensionality}. In this paper, a hybrid approach by combining CFS with BA is proposed to optimize the efficiency of the feature selection process and enhance the accuracy   of the classification. The main concept of this approach is to evaluate the relevance and the redundancy of the selected feature subset which is searched in the given search space for the optimal solution.

\subsubsection{Correlation-based feature selection (CFS)}
CFS~\cite{singh2018detection} is one of classical filter algorithms that choose features according to the result of the heuristic (correlation-based) assessment function. The preference of this function is to select subsets whose features are extraordinarily related with the class but uncorrelated with each other. While insignificant features that show low association with the class ought to be ignored on the grounds, repetitive features are chosen due to high relation with at least one of the rest of features. The acknowledgment of a feature will rely upon the degree to which it predicts classes in territories of the instance space not as of now anticipated by different features. The feature subset assessment function~\cite{singh2018web} in CFS is as:
\begin{equation}\label{equation1}
M_s=\frac{k\overline{r_{cf}}}{\sqrt{{k+k(k-1)+\overline{r_{ff}}}}}
\end{equation}

In Eq.~\ref{equation1}, $M_s$ is the heuristic evaluation for a feature subset $s$ including $k$ features, $\overline{r_{cf}}$ is the mean correlation degree between features and the category label, and $\overline{r_{ff}}$ is the average inter-correlation degree among features. The evaluation of CFS is a method of correlation based on feature subsets. A bigger $\overline{r_{cf}}$ or smaller $\overline{r_{ff}}$ in acquired subsets by the method produce a higher evaluation value, and the set of features with the highest value found during the process is utilised to reduce the size of both the training and testing set.

\subsubsection{Bat algorithm (BA)}
The original bat algorithm was developed by Xin-She Yang in 2010~\cite{yang2010new,yang2014nature}. The main inspirations for these works were the echolocation behavior of microbats. As BA uses frequency tuning, it is, in fact, the first algorithm of its kind in the context of optimization and computational intelligence. Each bat flies randomly with a velocity $v^t_i$, a location $x^t_i$, and a frequency $f_i$ at iteration $t$, in a $d-$dimensional search or solution space. The location can be considered as a solution vector to a problem of interest. Among the $n$ bats in the population, the current best solution $x_*$ found so far can be archived during the iterative search process.

Defined by Yang~\cite{yang2013bat}, the updating rules for location $x^t_i$ and velocity $v^t_i$ at time step $t$ are given by
\begin{gather}
	f_i=f_{min}+(f_{max}-f_{min})\beta \label{equation2}\\
	v^t_i=v^{t-1}_i+(x^{t-1}_i-x_*)f_i \label{equation3}\\
    x^t_i=x^{t-1}_i+v^t_i \label{equation4}
\end{gather}
\noindent where $\beta \in$ [0,1] is a random vector drawn from a uniform distribution.

For the local search part, once a solution is selected among the current best solutions, a new solution for each bat is generated locally using random walk
\begin{equation}\label{equation5}
x_{new}=x_{old}+\varepsilon A^t
\end{equation}
\noindent where $\varepsilon$ is a random vector drawn from a uniform distribution in [-1,1] or a Gaussian distribution, while $A^t$ is the average loudness of all the bats at this time step.

In addition, the loudness $A^t_i$ and the rate $r^t_i$ of pulse emission have to be updated accordingly as the iterations proceed. The updating rules for them can be written as
\begin{gather}
	A^{t+1}_i=\alpha A^t_i \label{equation6}\\
	r^{t+1}_i=r^0_i(1-e^{-\gamma t}) \label{equation7}
\end{gather}
\noindent where $0 <\alpha< 1$ and $\gamma> 0$ are constants.

\subsubsection{CFS-BA approach for feature selection}
In this section, we proposed CFS-BA based feature selection approach, which is used to assess the importance and the correlation of the selected feature subset. CFS-BA approach utilises correlation based feature technique to form the fitness functions and evaluation of integrity of the reduced feature subset.

For a feature subset $S$ with $k$ features, $S=(s_1,s_2,...s_k)$, CFS assesses the mean feature-class correlation and average inter-correlation among features by using Eq.~\ref{equation1}. As one of classical filter algorithms, CFS can easily select the subset of independently good features according to the result of correlation-based evaluation function. However, this feature subset may not be the best combination because of redundancy between features.

In order to remove the redundant features and reduce the dimensionality, BA, which inspired by the echolocation behavior of microbats, is introduced. In BA, every solution of the problem is denoted by the location of a bat, which can be represented by a vector. Bats fly in the search space to search for the best solutions and during this movement, the current best solution found so far can be archived. The population scans for the ideal arrangement by refreshing and updating the position of every bat based on Eq.~\ref{equation2}$-$Eq.~\ref{equation4} during the iterative search process.

The feature selection process of the CFS-BA approach is presented in Algorithm~\ref{algorithm1}. The main parts of the CFS-BA algorithm can be summarized as follows:

$\bullet$ Initialization (lines 1-4). The parameters of algorithm, generation and evaluation of the initial population are initialized here.

$\bullet$ New solution generation (lines 7-8). Here, bats in the population are moved in the search space according to updating rules of Eq.~\ref{equation2}$-$Eq.~\ref{equation4}.

$\bullet$ Local search process (lines 9-11). We select a solution among the best solutions, then generate a local solution around the selected one by random walks.

$\bullet$ Evaluation of the new solution (line 13). The feature subset assessment function in CFS is utilized here to evaluate the new solution.

$\bullet$ Archive of the new solution (line 14-17). The new solution which meets our requirement needs to be archived here. After that, the loudness $A^t_i$ and the rate $r^t_i$ of pulse emission have to be updated using Eq.~\ref{equation6}$-$Eq.~\ref{equation7}.

$\bullet$ Update of the best solution (line 19-20). We compare the evaluation result of the archived solution and find the current best $X_{best}$ until the iterations end.

\begin{algorithm}[t]
  \caption{ CFS-BA approach for feature selection}
  \begin{algorithmic}[1]\label{algorithm1}
    \REQUIRE ~~Training Dataset and Testing Dataset\\
    \ENSURE ~~Selected Feature Subset $X_{best}$\\
    \STATE Initialize a population of $n$ bats $X_i=(x_{i1},...x_{iD})^T(i=1,2,...,n)$ and $v_i$\\
    \STATE Initialize frequency $f_i$, pulse emission rate $r_i^t$, and loudness $A_i^t$\\
    \STATE Initialize $fit(X_i)$ (cf. Eq.1) and $X_{best}$\\
    \STATE Initialize $fit_{temp}(i)$ and $X_{temp}(i)$ for solution storage\\
    \WHILE{$1 \leqslant t \leqslant$ Max no. of iterations}
    \FOR{$i=1$ to $n$}
       \STATE Generate new $f_i$ (cf. Eq.2)
       \STATE Update $X_i$ and $v_i$ (cf. Eq.3 and Eq.4)
       \IF{$r_i^t <$  rand(0,1)}
        \STATE Select a $X_i$ from $X_{best}$
        \STATE Generate a new $X_{new}$ (cf. Eq.5)
       \ENDIF
       \STATE Calculate $fit(X_{new})$ (cf. Eq.1)
       \IF{$fit(X_i) \leqslant fit(X_{new})$ and N(0,1) $< A_i^t$}
        \STATE $fit_{temp}(i) \gets fit(X_{new})$
        \STATE $X_{temp}(i) \gets X_{new}$
        \STATE Decrease $A_i^t$ and Increase $r_i^t$ (cf. Eq.6 and Eq.7)
       \ENDIF
       \IF{$fit(X_{new}) \geq$ Max of $fit_{temp}$}
        \STATE $X_{best} \gets X_{new}$
       \ENDIF
    \ENDFOR
        \STATE $t=t+1$
    \ENDWHILE
  \end{algorithmic}
\end{algorithm}

\subsection{Ensemble classification}\label{subsection2}
For ensemble learning, the classification methods usually combine multiple base classifiers in some way to produce better accuracy~\cite{feng2018dynamic}. These classifiers are powerful to solve the same problem and collectively achieve a forecasting result with higher stability and accuracy by creating multiple independent models and combining them~\cite{li2013predicting}. The classical reasons for employing ensemble classifiers to improve the effectiveness are representational issue, statistical reason, and computational reason. First, sometimes a single classifier is not qualified to obtain the best representation in the hypothesis space, therefore, it is necessary to combine independent classifiers to improve the predictive performance. Second, if the input dataset is not sufficient to train the learning algorithm, the result may lead to a weak or false hypothesis. In the last case, in order to produce a suitable hypothesis, an individual classifier could spend a significant amount of computing time, in which the procedure will be more likely to cause problems.

Bagging~\cite{breiman1996bagging} and Boosting~\cite{freund1996experiments} are the two most popular algorithms in ensemble learning, usually producing good results in classification and being widely chosen to build many ensemble models. Moreover, the other well-known ensemble learning methods for improving the performance of classification are Voting~\cite{hu2018approach} , Bayesian parameter averaging~\cite{friston2010generalised}, and Stacking~\cite{hung2009selective}. Likewise, ensemble methods have been shown to improve accuracy in many use cases, including intrusion detection. For example, the results in~\cite{abdullah2018enhanced,pham2018improving,salo2019dimensionality} proved that their proposed ensemble models produce better performance of IDS than the one using a single classifier. For security professionals, ensemble classifiers provide mechanisms that aid in analysis such as similarity to existing known malicious or benign samples.

Among decision tree algorithms, C4.5 has been widely used in the field of anomaly detection due to its high  efficiency and its simple characteristics. Meanwhile, random forest is the most representative algorithm among ensemble learning methods, and it is generally more robust and can achieve better performances than single decision trees. Moreover, Forest PA can use the strength of the entire feature space to generate trees with high accuracy. With its novel weight assignment strategy and bootstrap sampling, Forest PA generates highly diverse trees while retaining their higher individual accuracy. Therefore, C4.5, random forest, and Forest PA are selected to construct the ensemble for multi-class intrusion detection in this paper.

For bagging algorithm, the base classifiers are generated in parallel by bootstrap sampling. Boosting works by training a set of classifiers sequentially and combining them for prediction, where the later classifiers focus more on the mistakes of the earlier classifiers. However, sensitivity to noise leads to performance degradation when appearing wrong labels. Moreover, base classifiers usually are homogeneous in bagging and boosting, which will be not suitable for three different base classifiers (C4.5, random forest, and Forest PA) in this paper. Although stacking generates an ensemble of heterogeneous learners, it will bring enormous computational complexity when generating different level models. Compared to the above algorithms, in this paper, voting is more suitable for heterogeneous learners ensemble with lower computational complexity and less time overhead.

\subsubsection{C4.5}
C4.5~\cite{quinlan2014c4} is a typical decision tree algorithm which is developed based on the ID3~\cite{hssina2014comparative} algorithm. This algorithm passes through decision tree, visits each node and select optimal split based on the maximisation of the gain ratio, which is represented by the following formula:
\begin{equation}
GainRatio(A)=\frac{Gain(A)}{SplitInfo(A)}
\end{equation}

In the process, an attribute with the highest information gain is chosen as splitting attribute for the node $N$. Information gain represents how much uncertainty in the set $D$ is reduced after it is partitioned on attribute $A$, where the uncertainty can be calculated by entropy as:
\begin{equation}
Entropy(D)=-\sum\limits_{x \in X} p(x)log_2 p(x)
\end{equation}
\noindent where $X$ is the set of classes in $D$ and $p(x)$ is the proportion of number of elements in class $x$ to the number of elements in set $D$.

Likewise, $SplitInfo$ is the term which describes how equally the attribute splits the data and can be calculated as:
\begin{equation}
SplitInfo(A)=-\sum\limits_{j=1}^n \frac{|D_j|}{|D|}log_2\left(\frac{|D_j|}{|D|}\right)
\end{equation}
\noindent where $\frac{|D_j|}{|D|}$ represents the weight of the $j-$th partition in the set $D$.

Moreover, as an improvement of ID3 algorithm, C4.5 has the capability to model or classify both discrete and continuous attributes, and can ignore missing attribute values in a dataset.

\subsubsection{Random Forest (RF)}
Random Forest, proposed by Breimanis in~\cite{breiman2001random}, is another decision tree technique that operates by constructing multiple decision trees. It takes thousands of input variables without variable deletion and classifies them based on their significance. RF can be described as an ensemble of classification trees where every tree contributes with a single vote for the task of the most frequent class to the input data. Compared to other machine learning methods (e.g., support vector machine, artificial neural network), there are fewer parameters to be specified when running RF. In RF, a collection of individual tree structured classifiers can be defined as:
\begin{equation}
\left\{h\left(x,\theta_k\right),k=1,2,...i...\right\}
\end{equation}

\noindent where $h$ represents RF classifier, $\left\{\theta_k\right\}$ stands for random vectors distributed independently identical, and each tree has a vote for the most famous class at input variable $x$. The nature and dimensionality of $\theta$ depends on its use in tree construction.

The key to the success of RF is the creation of each decision tree that makes up the forest. A bootstrapped subset of the training dataset is created to train each tree in the forest. Due to this fact, on average, each tree makes use of around two-thirds of the training dataset. The unused elements are called by the Out Of Bag (OOB) samples, which are used for inner cross-validation to evaluate the classification accuracy of RF.

Significantly, RF has a low computational burden, and it is insensitive to the parameters and outliers. Besides, over-fitting is less of an issue compared to individual decision tree, and there is no need to prune the trees which is a cumbersome task~\cite{feng2015uav}.

\subsubsection{Forest by Penalizing Attributes (Forest PA)}
Unlike some existing algorithms that use a subset of the non-class attributes, Forest PA~\cite{adnan2017forest} is an algorithm that builds a set of highly accurate decision trees by exploiting the strength of all non-class attributes available in a data set. At the same time, some weight-related concerns, such as weight assignment strategy and weight increment strategy, are taken into account in order to retain individually accurate and promote strong diversity.

For the weights of the attributes that appear in the latest tree, Forest PA will randomly update the weights for those attributes within a Weight-Range (WR), which can be defined as follows:
\begin{equation}
WR^{\lambda}=\left\{
\begin{aligned}
 & [0.0000,e^{-\frac{1}{\lambda}}], &\lambda=1 \\
 & [e^{-\frac{1}{\lambda-1}}+\rho,e^{-\frac{1}{\lambda}}], &\lambda>1
\end{aligned}
\right.
\end{equation}
\noindent where $\lambda$ represents the level of the attribute and $\rho$ is used to ensure the WR for different levels be non-overlapping. For example, if an attribute appears in the root node then its $\lambda=1$. In the same way, if an attribute is tested at a child node of the root node then its $\lambda=2$.

Moreover, in order to address the negative effect of retaining weights which are not present in the latest tree, Forest PA has a mechanism to gradually increase weights of the attributes that have not been tested in the subsequent trees. Let an attribute $A_i$ is tested at Level $\rho$ of the $T_{j-1}-$th tree with $\eta$ height and its weight is $\omega_i$. Then, the weight increment value $\sigma_i$ of $A_i$ is calculated as:
\begin{equation}
\sigma_i=\frac{1.0-\omega_i}{(\eta+1)-\lambda}
\end{equation}

\subsubsection{Vote}\label{subsection_vote}
Vote is a meta algorithm which performs the decision process by applying several classifiers~\cite{catal2017sentiment}. It uses the power of several individual classifiers and applies a combination rule for the decision. For example, minimum probability, maximum probability, majority voting, product of probabilities, and average of probabilities are different algorithms for combination rules. In order to deal with the multi-class classification, majority voting could not be chosen because the number of classes is more than that of base classifiers. In this paper, average of probabilities approach is used to make decision, where the class label is determined based on the maximum value of the average of predicted probabilities.

Suppose we have $l$ classifiers $C = \left\{C_1,...,C_l\right\}$, and $c$ classes $\Omega = \left\{\omega_1,...,\omega_c\right\}$. For instance, due to the above base classifiers considered in our experiment, $l$ can be set to $3$, and the value of $c$ depends on the number of attack types. A classifier $C_i:R^n \to [0,1]^c$ accepts an object $x \in R^n$ and outputs a vector $\left[P_{c_i}(\omega_1|x),...P_{c_i}(\omega_c|x)\right]$, where $P_{c_i}(\omega_j|x)$ denotes the probability assigned by the classifier $C_i$ that object $x$ belongs to class $\omega_j$. For each class $\omega_j$, let $m_j$ represents the mean of the probabilities assigned by the $l$ classifiers, which can be calculated as:
\begin{equation}
m_j=\frac{1}{l}\sum\limits_{i=1}^l P_{c_i}(\omega_j|x)
\end{equation}

let $M=\left[m_1,...m_c\right]$ be the set of mean probabilities for $c$ classes. Then, $x$ is assigned to the class $\omega_k$ if $m_k$ is the maximum in $M$.

\section{Evaluations and results}\label{section4}
As stated before, this paper aims to develop an efficient intrusion detection system with high accuracy and low false alarms. For this purpose, a hybrid method, combined CFS and BA named CFS-BA, is performed to determine a subset of the original features in order to eliminate the irrelevant features, and improve the classification efficiency. In the classification step, an ensemble classifier combined three different algorithms, C4.5, RF, and Forest PA based on the AOP combination rule, is trained and tested based on three datasets. The experiments are performed by Weka 3.8.3~\cite{witten2016data} on desktop PC with 3.6 GHz Intel Core i7-4790 processor and 16GB RAM.
\subsection{Description of the benchmark datasets}\label{subsection4.1}
During the evaluation of IDS, one of the challenges faced by researchers is finding a suitable dataset. Acquiring a real world dataset that represents the traffic flowing through the network without any sort of anonymization or modification is a problem that has been continuously encountered by the cybersecurity research community~\cite{aldwairi2018evaluation}. Even in the cases where the data is allowed to be released or shared for public use, it will be heavily anonymized or severely altered. This will cause a lot of the essential data components that are considered critical to the researchers to be lost or no longer reliable.

For this reason, many researchers have decided to use simulated datasets such as the most well-known KDDCup'99 dataset~\cite{rosset2000kdd}, or one of its contemporaries the NSL-KDD dataset~\cite{tavallaee2009detailed}. Recently there has been a significant effort to try and develop data sets that are reflective of real world data. In 2015, Kolias et al.~\cite{kolias2015intrusion} published Aegean WiFi Intrusion Dataset (AWID) dataset, which includes real traces of both normal and intrusive 802.11 traffic. In addition, in 2017, the Canadian Institute for Cybersecurity (CIC) published an intrusion detection dataset named CIC-IDS2017~\cite{sharafaldin2018toward}, which resembles the true real-world data packet capture (PCAPs). Therefore, in this paper, experiments are conducted based on the NSL-KDD, AWID, and CIC-IDS2017 datasets.

\subsubsection{NSL-KDD dataset}
The NSL-KDD dataset~\cite{tavallaee2009detailed} was proposed in 2009 as a new revised version of the original dataset KDDCup'99~\cite{lee1999data}. On the one hand, NSL-KDD retained the advantageous and challenging characteristics of KDDCup'99. On the other hand, it addressed some drawbacks inherited from the original dataset by eliminating redundant records, rationalizing the number of instances, and maintaining the diversity of selected samples. It is worth noting that the NSL-KDD dataset is compiled to maximize the difficulty of prediction, which constitutes its outstanding characteristics. In order to group the records into five difficulty levels, the initial dataset was evaluated using several benchmark classifiers, and each instance was annotated with the number of its successful predictions~\cite{bala2019review}. For each difficult level group, the amount of selected records is inversely proportional to the record percentages from the original KDDCup'99 dataset.

In this study, KDDTrain+, KDDTest+, and KDDTest-21 sets of the NSL-KDD dataset are used. The KDDTrain+ set contains total 125,973 instances comprising of 58,630 instances of attack traffic and 67,343 instances of normal traffic. Whereas, the KDDTest+ set contains total 22,544 instances, and as a subset of the KDDTest+ set, the KDDTest-21 set includes total 11,850 instances. Cross-validation is done on the the KDDTrain+ set in our experiments, and to extend this benchmark, we also consider a validation test using simple hold-out (train-test) approach applied on KDDTest+ and KDDTest-21 sets. A detailed overview of the instances is shown in Table~\ref{table1}.

\subsubsection{Aegean WiFi Intrusion Dataset (AWID) dataset}

AWID was publicly available in 2015 as a collection of sets of WiFi network data, which contain real traces of both normal and intrusive data collected from real network environments~\cite{kolias2015intrusion}. Each record in the dataset is represented as a vector of 155 attributes, and each attribute has numeric or nominal values. Based on the number of target classes, the dataset can be classified into AWID-CLS dataset and AWID-ATK dataset. AWID-CLS dataset groups the instances into 4 main classes including normal, flooding, impersonation, and injection, while AWID-ATK dataset has 17 target classes that belong to the 4 main classes. On the other hand, based on the number of instances, all the datasets have two different versions: Full Set and Reduced Set. It is important to mention that these two versions are not related. The reduced set was collected independently from the full set at different times, with different tools, and in different environments.

For this research we have conducted experiments on the the reduced four class dataset (AWID-CLS-R-Tst) by using cross-validation method for classification purposes. In general, AWID-CLS-R-Tst set includes total 575,643 instances, and more detailed information about the numbers of specific attacks can be seen in Table~\ref{table2}.

\subsubsection{CIC-IDS2017 dataset}

The CIC-IDS2017 dataset was published by Canadian Institute for Cybersecurity (CIC) in 2017, it contains benign and the most up-to-date common attacks~\cite{sharafaldin2018toward}. It also includes the results of the network traffic analysis using CICFlowMeter with labeled flows based on the time stamp, source and destination IPs, source and destination ports, protocols, and attacks (CSV files). This is one of the newest intrusion detection datasets, which covers necessary criteria with updated attacks such as DDoS, Brute Force, XSS, SQL Injection, Infiltration, Port Scan, and Botnet. In detail, this dataset contains 2,830,743 records devised on 8 files and each record includes 78 different features with its label.

In order to maintain the same order of magnitude of each dataset while taking into account the requirements of multi-classification, the Wednesday-workingHours set has been chosen for experiments through cross-validation method. This set includes total 691,406 instances belonging to 6 categories, and the static information of the set is given in Table~\ref{table2}.

\begin{table}[pos=t]
\caption{Statistics of the three sets of the NSL-KDD dataset.}
\label{table1}
\begin{tabular}{@{}llllll@{}}
\toprule
Class         & \multicolumn{5}{c}{NSL-KDD}             \\ \cmidrule(l){2-6}
              & KDDTrain+ &  & KDDTest+ &  & KDDTest-21 \\ \midrule
Normal        & 67343     &  & 9711     &  & 2152       \\
DoS           & 45927     &  & 7458     &  & 4342       \\
PRB           & 11656     &  & 2421     &  & 2402       \\
R2L           & 995       &  & 2754     &  & 2754       \\
U2R           & 52        &  & 200      &  & 200        \\ \midrule
Attacks & 58630     &  & 12833    &  & 9698       \\ \midrule
Total         & 125973    &  & 22544    &  & 11850      \\ \bottomrule
\end{tabular}
\end{table}

\begin{table}[pos=t]
\caption{Statistics of the AWID and CIC-IDS2017 datasets.}
\label{table2}
\begin{tabular}{@{}llll@{}}
\toprule
Class         & AWID      & Class            & CICIDS-2017 \\ \cmidrule(lr){2-2} \cmidrule(l){4-4}
              & CLS-R-Tst &                  & Wed.        \\ \midrule
Normal        & 530785    & Normal           & 439683      \\
Injection     & 16682     & DoS slowloris    & 5796        \\
Flooding      & 8097      & DoS Slowhttptest & 5499        \\
Impersonation & 20079     & Dos Hulk         & 230124      \\
              &           & DoS GoldenEye    & 10293       \\
              &           & Heartbleed       & 11          \\ \midrule
Attacks       & 44858     & Attacks          & 251723      \\ \midrule
Total         & 575643    & Total            & 691406      \\ \bottomrule
\end{tabular}
\end{table}

\subsection{Dataset preprocessing}

Data preprocessing is the most time consuming and essential step in data mining. Realistic data typically comes from heterogeneous platforms and can be noisy, redundant, incomplete, and inconsistent~\cite{li2018feature}. Thus, it is important to transform raw data into a format suitable for analysis and knowledge discovery. Therefore, in this research, the preprocessing step involves data filtration, data transforming as well as data normalization.
\subsubsection{Data filtration}
Due to the heterogeneity of the platforms, the raw data inevitably contain anomalous and redundant instances, which may have a negative influence on classification accuracy. In order to solve this problem, these records need to be removed from the dataset at the beginning of our experiments. For instance, the feature `Fwd Header Length' appears twice in the CIC-IDS2017 dataset, and `Flow Packets/s' includes abnormal values such as `Infinity' and `NaN'. Moreover, we have replaced missing values with zeroes and dropped out the features with constants values as they do not contribute to the class distinction. For example, the AWID-CLS-R-Tst set remains 84 features from the original 155 ones after data filtration.

\subsubsection{Data transforming and normalization}
The utilized datasets contain symbolic, continuous, and binary values. For instance, the feature `protocol type' in the NSL-KDD datasets includes symbolic values such as: `tcp', `udp', and `icmp'. As many classifiers accept only numerical values, the converting process is considered vital and has a significant impact on IDS accuracy. In this paper, we replace every single value with an integer in order to handle the symbolic features. Moreover, different scales among features can degrade the classification performance, for example, features that take on large numeric values, e.g., for the CIC-IDS2017 dataset, `Flow Duration' can dominate the classifier's model relative to features with relatively small numeric values such as `Total Fwd Packets'. Accordingly, normalization is a `scaling down' transformation which maps features onto a normalized range. A simple and fast approach called minimum-maximum method~\cite{kotsiantis2006data} is used in our experiments, which can be defined as:
\begin{equation}
\overline{x}=\frac{x-x_{min}}{x_{max}-x_{min}}
\end{equation}
\noindent where $x_{min}$ and $x_{max}$ represent the minimum and maximum values of feature $x$.

\subsection{Results and discussion}
\begin{table*}[]
\caption{Selected features for the NSL-KDD, AWID, and CIC-IDS2017 datasets.}
\centering
\label{table3}
\begin{tabular}{@{}llllllll@{}}
\toprule
\multicolumn{2}{c}{NSL-KDD}            &  & \multicolumn{2}{c}{AWID}        &  & \multicolumn{2}{c}{CIC-IDS2017}   \\ \cmidrule(r){1-2} \cmidrule(lr){4-5} \cmidrule(l){7-8}
No. & Feature Name                     &  & No. & Feature Name              &  & No. & Feature Name                \\ \cmidrule(r){1-2} \cmidrule(lr){4-5} \cmidrule(l){7-8}
3   & service                          &  & 1   & frame.time\_epoch         &  & 1   & Destination Port            \\
4   & flag                             &  & 15  & radiotap.datarate         &  & 6   & Total Length of Bwd Packets \\
5   & src\_bytes                       &  & 16  & radiotap.channel.freq     &  & 13  & Bwd Packet Length Mean     \\
6   & dst\_bytes                       &  & 17  & radiotap.channel.type.cck &  & 15  & Flow Bytes/s               \\
14  & root\_shell                      &  & 24  & wlan.fc.frag              &  & 17  & Flow IAT Mean              \\
26  & srv\_serror\_rate                &  & 29  & wlan.duration             &  & 34  & Bwd Header Length           \\
29  & same\_srv\_rate                  &  & 32  & wlan.ta                   &  & 37  & Min Packet Length           \\
30  & diff\_srv\_rate                  &  & 35  & wlan.frag                 &  & 50  & Down/Up Ratio               \\
37  & dst\_host\_srv\_diff\_host\_rate &  &   &                             &  & 57  & Subflow Bwd Bytes           \\
39  & dst\_host\_srv\_serror\_rate     &  &   &                             &  & 58  & Init\_Win\_bytes\_forward    \\
&                                      &  &   &                             &  & 59  & Init\_Win\_bytes\_backward  \\
&                                      &  &   &                             &  & 67  & Idle Std                    \\
&                                      &  &   &                             &  & 68  & Idle Max                    \\ \bottomrule
\end{tabular}
\end{table*}

\begin{figure*}
\centering
  \subfigure[NSL-KDD]{\includegraphics[width=2.1in,height=1.9in]{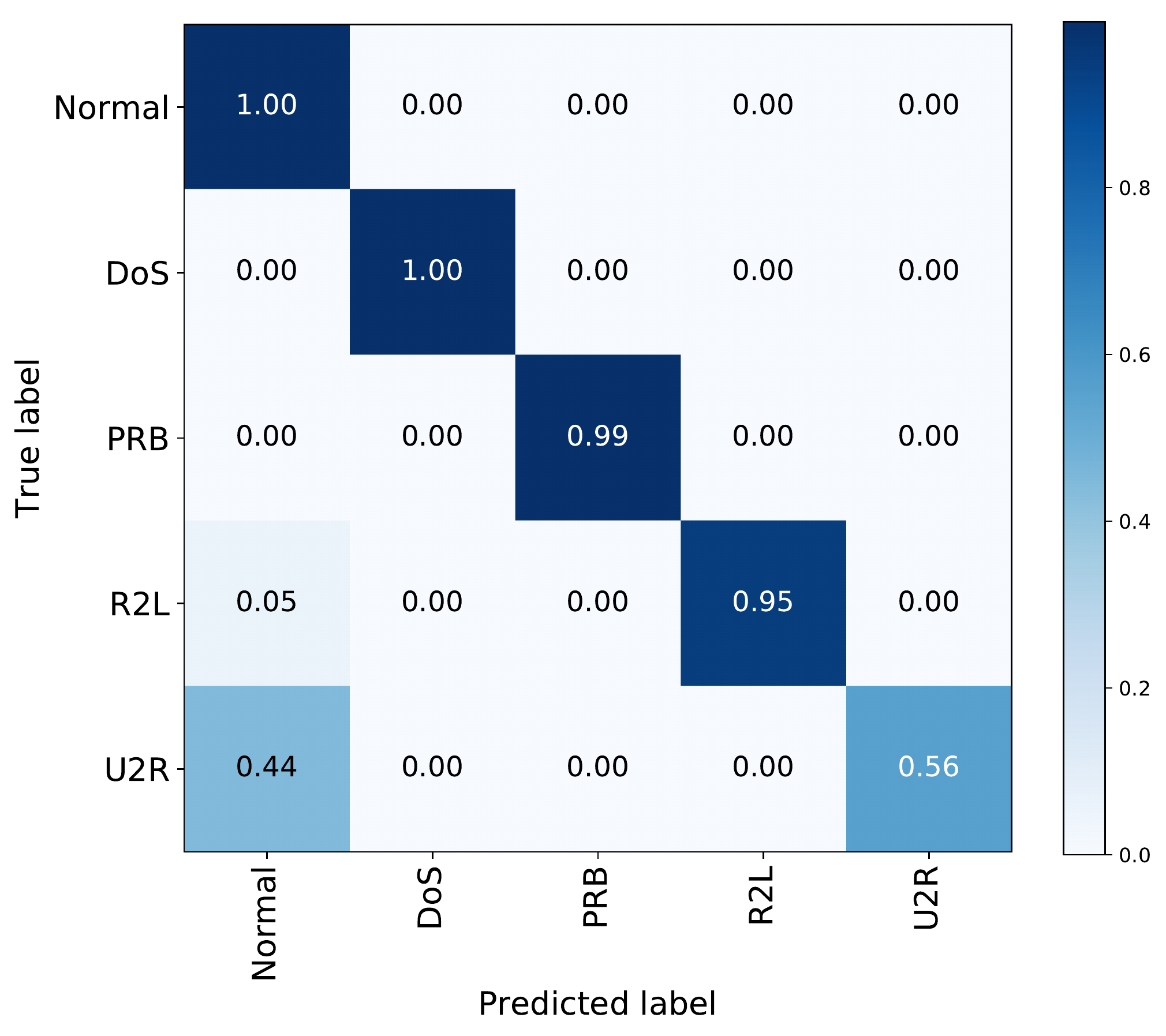}
  \label{fig2(a)}}
  \quad
  \subfigure[AWID]{\includegraphics[width=2.1in,height=1.9in]{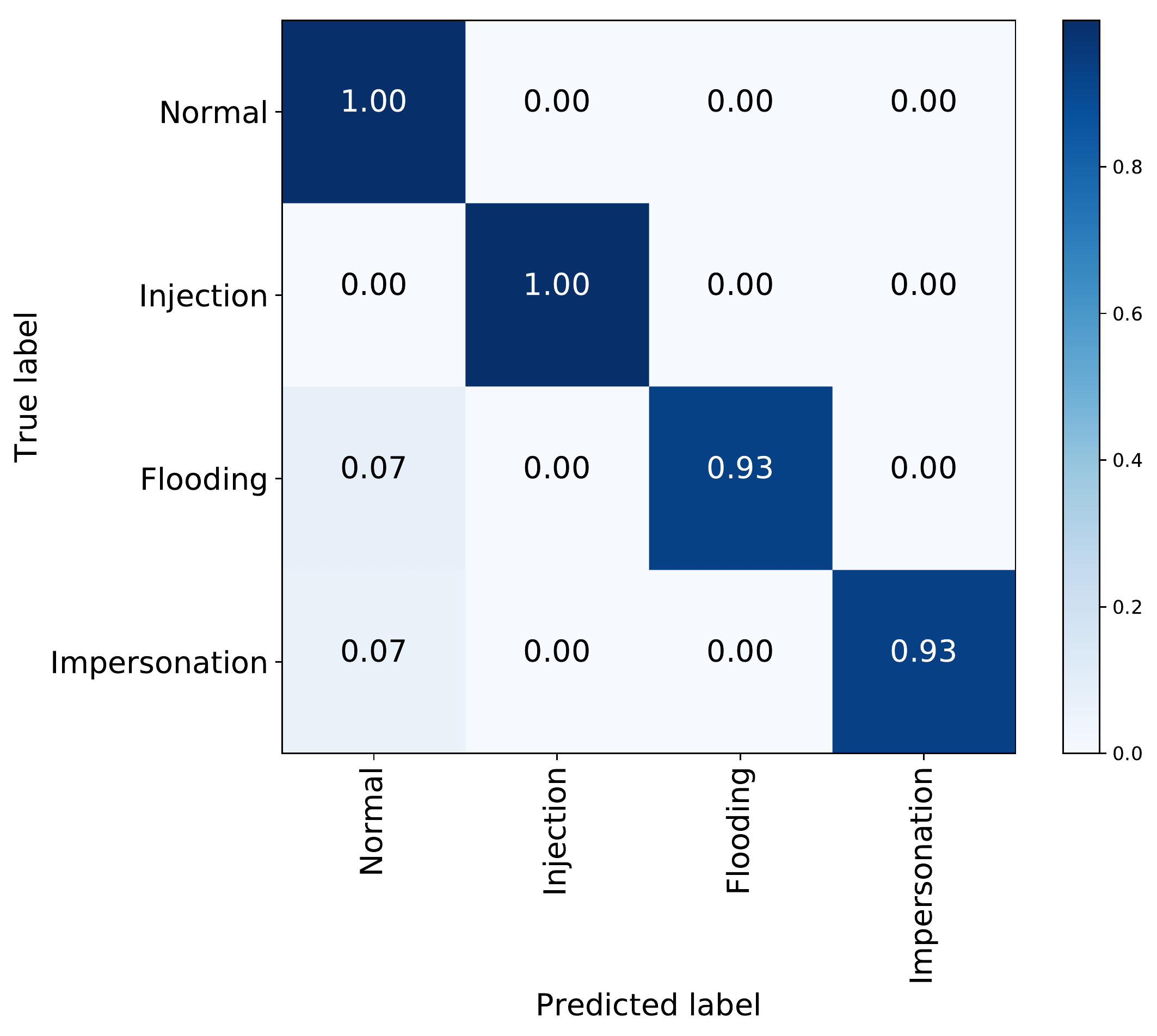}
  \label{fig2(b)}}
  \quad
  \subfigure[CIC-IDS2017]{\includegraphics[width=2.1in,height=1.9in]{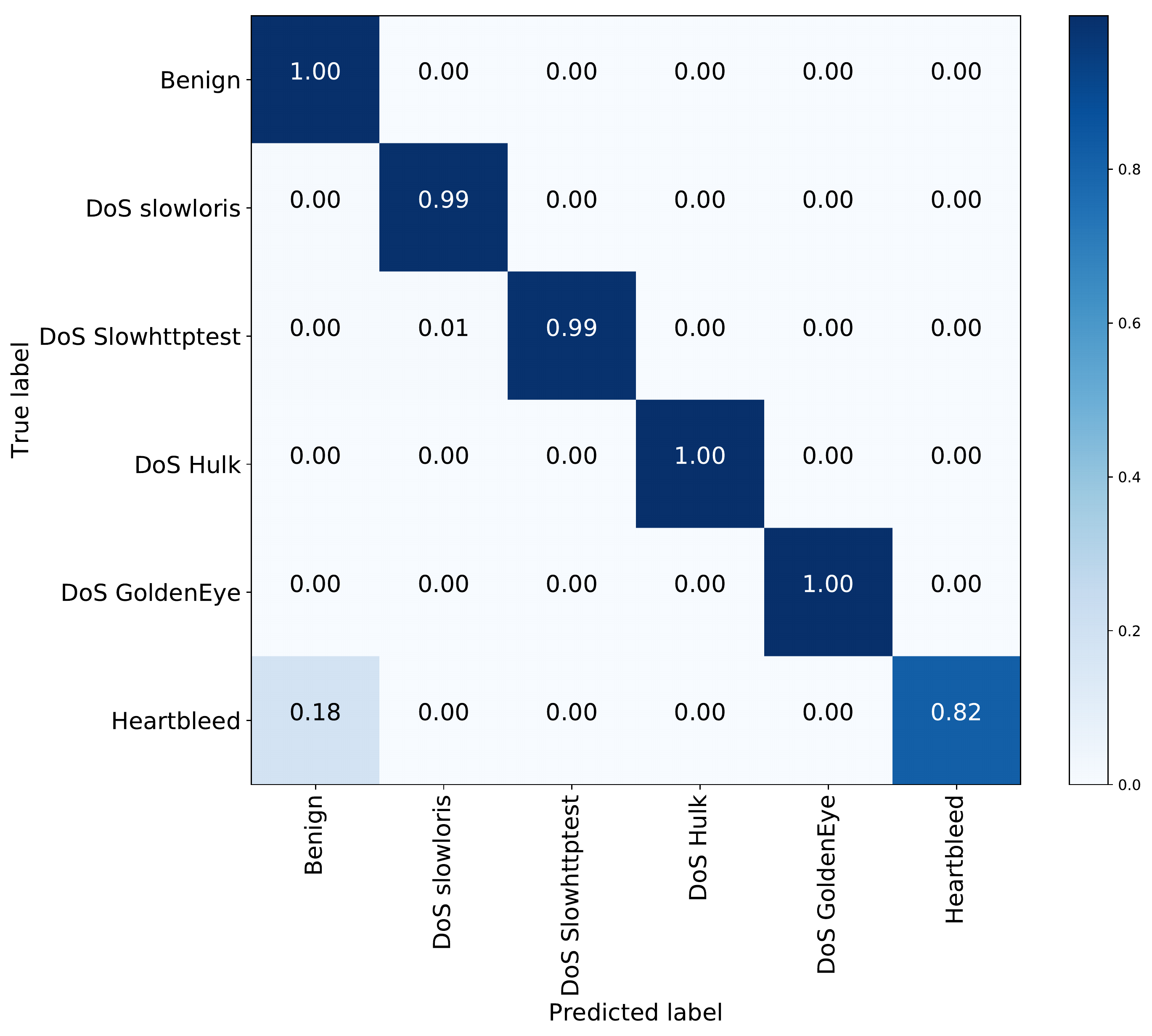}
  \label{fig2(c)}}

\caption{Normalized confusion matrices based on the NSL-KDD, AWID, and CIC-IDS2017 datasets.}
\label{fig2}
\end{figure*}

The performance of IDS is evaluated based on its capability of classifying network traffic into a correct type. In order to avoid the effect of data sampling when assessing the IDS, therefore, we conducted experiments by using repeated k-fold (kf) cross-validation method, and the value of k is considered as 10. In this paper, all the performance results reported are the average value of outputs from 10 iterations of 10f validation approach, and each experiment is repeated with different seed for avoiding biased results. More specifically, for each dataset, we provide the confusion matrix derived from the testing process of CFS-BA-Ensemble, and compare the performance of the proposed algorithm with no feature selection and some state-of-the-art methods in terms of several detection metrics, including Accuracy (Acc), precision, Detection Rate (DR), F-Measure, Attack Detection Rate (ADR), and False Alarm Rate (FAR). The mathematical calculations of the utilized evaluation metrics are explained in~\cite{elhag2019multi}.

First, essential features are identified by utilizing the proposed CFS-BA approach to evaluate the integrity of the reduced feature subset in the feature selection stage. Then, candidate features are selected from the original ones for the next stage. Table~\ref{table3} shows the numbers and names of selected features for NSL-KDD, AWID, and CIC-IDS2017 datasets. By implementing CFS-BA alone, the approach is seen to reduce the dimensionality drastically and eliminate the irrelevant features of the dataset. Finally, in order to significantly improve the predictive performance of IDS, an ensemble classifier which consists of three different decision tree classifiers is used in a vote algorithm.

\begin{table*}[]
\caption{Performance classification for feature selection based on NSL-KDD with 10f validation.}
\centering
\label{table-nsl-kdd}
\begin{tabular}{llllllll}
\hline
\multicolumn{8}{l}{(a).The performance results based on the original features (41 features)}                              \\ \hline
Classifier & Acc            & Precision      & DR             & F-Measure      & ADR            & FAR            & MBT(s) \\ \hline
C4.5       & 0.941          & 0.945          & 0.941          & 0.943          & 0.913          & 0.035          & 16.91  \\
RF         & 0.949          & 0.944          & 0.949          & 0.947          & 0.903          & 0.021          & 14.98  \\
ForestPA   & 0.945          & 0.942          & 0.945          & 0.944          & 0.913          & 0.028          & 43.16 \\
Ensemble   & 0.953          & 0.951          & 0.953          & 0.952          & 0.919          & 0.016          & 51.44 \\
\multicolumn{8}{l}{(b).The performance results based on the selected features using CFS-BA (10 features)}                 \\
Classifier & Acc            & Precision      & DR             & F-Measure      & ADR            & FAR            & MBT(s) \\
C4.5       & 0.988          & 0.987          & 0.988          & 0.988          & 0.986          & 0.012          & 2.93   \\
RF         & 0.991          & 0.988          & 0.991          & 0.989          & 0.987          & 0.009          &  8.63  \\
ForestPA   & 0.987          & 0.989          & 0.987          & 0.988          & 0.985          & 0.008          & 29.66  \\
Ensemble   & \textbf{0.998} & \textbf{0.998} & \textbf{0.998} & \textbf{0.998} & \textbf{0.997} & \textbf{0.001} & 36.28  \\ \hline
\end{tabular}
\end{table*}

Fig.~\ref{fig2(a)}, Fig.~\ref{fig2(b)}, and Fig.~\ref{fig2(c)} separately indicate the multi-class classification performance of the proposed IDS with 10f cross-validation among the NSL-KDD, AWID, and CIC-IDS2017 datasets. It is observed that the performance of most classifications is adequate, while several attacks can not be classified very well, such as `U2R' and `Heartbleed'. As seen in Table~\ref{table1} and Table~\ref{table2}, the numbers of these instances are much less than others, which significantly affects the classification results of these attacks. In detail, there are only 11 instances with the label `Heartbleed' out of 251,723 attack instances in the CIC-IDS2017 and 52 `U2R' instances in the KDDTrain+ set, which poses a challenge for the IDS to correctly classify them. In general, the proposed method is not focused on a specific class, it is proposed for selecting relevant features for all classes, which could not guarantee the performance of every type of attacks, especially some attacks with very few instances in the datasets. However, as the classification results for normal instances are pretty well among these datasets, the developed system can be used for intrusion detection.

\subsubsection{Comparison with no feature selection}

In order to evaluate the performance of the proposed IDS, we make a comparison between the proposed feature selection approach and without feature selection to distinguish attacks from benign instances. Thanks to the selection of relevant features by the proposed CFS-BA algorithm, the average values of these metrics, such as Acc, precision, DR, F-Measure, and ADR, have increased significantly.

Table~\ref{table-nsl-kdd} summarizes the performance based on the NSL-KDD dataset, which includes the results of the base and ensemble classifiers. It is indicated that the ensemble classifier is not good enough in some metrics without implementing feature selection. By contrast, the proposed CFS-BA-Ensemble method performs best on all the three sets. In detail, our model exhibits the highest accuracy of 0.998, F-Measure of 0.998, ADR of 0.997 and the lowest FAR of 0.001 based on the NSL-KDD dataset. As seen in Table~\ref{table-awid}, the proposed CFS-BA-Ensemble approach still achieves the best performance results in most respects on the AWID dataset, such as the highest accuracy of 0.995, the highest ADR of 0.959, and the lowest FAR of 0.002. Each base classifier using the selected feature exhibits higher accuracy and ADR than the ensemble classifier with the original features, which strongly proves the effectiveness of the proposed feature selection method. Similarly, the result of the comparison on the CIC-IDS2017 dataset is shown in Table~\ref{table-cic}, we observe that the performance of the proposed feature selection approach outperforms that of all features in every respect, and the CFS-BA-Ensemble approach achieves the highest accuracy rate of 0.999, DR of 0.999, and ADR of 0.999 with only 13 features, which also outperforms all other individual classifiers. In contrast, the best accuracy values of the C4.5, RF, and ForestPA classifiers are 0.983, 0.993, 0.988 using CFS-BA based feature selection method, respectively.

\begin{table*}
\caption{Performance classification for feature selection based on AWID with 10f validation.}
\centering
\label{table-awid}
\begin{tabular}{llllllll}
\hline
\multicolumn{8}{l}{(a).The performance results based on the original features (84 features)}                     \\ \hline
Classifier & Acc            & Precision      & DR    & F-Measure      & ADR            & FAR            & MBT(s) \\ \hline
C4.5       & 0.954          & 0.953          & 0.999 & 0.976          & 0.789          & 0.034          & 94.93  \\
RF         & 0.979          & 0.982          & 0.996 & 0.989          & 0.783          & 0.004          & 142.84 \\
ForestPA   & 0.966          & 0.982          & 0.981 & 0.981          & 0.784          & 0.019          & 435.11 \\
Ensemble   & 0.982          & 0.982          & 0.999 & 0.990          & 0.784          & 0.002          & 488.46 \\
\multicolumn{8}{l}{(b).The performance results based on the selected features using CFS-BA (8 features)}         \\
Classifier & Acc            & Precision      & DR    & F-Measure      & ADR            & FAR            & MBT(s) \\
C4.5       & 0.985          & 0.985          & 0.985 & 0.985          & 0.913          & 0.010          & 9.96   \\
RF         & 0.992          & 0.992          & 0.992 & 0.992          & 0.945          & 0.004          & 26.51  \\
ForestPA   & 0.990          & 0.989          & 0.990 & 0.989          & 0.902          & 0.003          & 79.93  \\
Ensemble   & \textbf{0.995} & \textbf{0.995} & 0.995 & \textbf{0.995} & \textbf{0.956} & \textbf{0.001} & 92.62  \\ \hline
\end{tabular}
\end{table*}

\begin{table*}
\caption{Performance classification for feature selection based on CIC-IDS2017 with 10f validation.}
\centering
\label{table-cic}
\begin{tabular}{llllllll}
\hline
\multicolumn{8}{l}{(a).The performance results based on the original features (78 features)}              \\ \hline
Classifier     & Acc       & Precision     & DR        & F-Measure     & ADR      & FAR      & MBT(s)     \\ \hline
C4.5           & 0.960     & 0.961         & 0.984     & 0.973         & 0.918    & 0.016    & 212.59     \\
RF             & 0.968     & 0.985         & 0.981     & 0.983         & 0.946    & 0.019    & 244.85     \\
ForestPA       & 0.967     & 0.978         & 0.984     & 0.981         & 0.938    & 0.016    & 859.62    \\
Ensemble       & 0.977     & 0.991         & 0.988     & 0.990         & 0.956    & 0.012    & 977.94    \\
\multicolumn{8}{l}{(b).The performance results based on the selected features using CFS-BA (13 features)} \\
Classifier     & Acc       & Precision     & DR        & F-Measure     & ADR      & FAR      & MBT(s)     \\
C4.5           & 0.983     & 0.996         & 0.989     & 0.992         & 0.974    & 0.011    & 32.02      \\
RF             & 0.993     & 0.995         & 0.998     & 0.996         & 0.984    & 0.003    & 58.04     \\
ForestPA       & 0.988     & 0.993         & 0.988     & 0.991         & 0.978    & 0.006    & 80.82     \\
Ensemble       & \textbf{0.999}     & \textbf{0.999}         & \textbf{0.999}     & \textbf{0.999}         & \textbf{0.999}    & \textbf{0.001}    & 98.42     \\ \hline
\end{tabular}
\end{table*}

Furthermore, due to the dimensionality reduction of the subsets, the proposed CFS-BA-Ensemble model reduces the time overhead when it is applied to the feature selection and ensemble model. Table~\ref{table-nsl-kdd}-Table~\ref{table-cic} also show a comparison of the average model building time (MBT) consumed by single training based on the different numbers of features. For the NSL-KDD dataset, although it does not take much time to build the ensemble model on this set, the reduction still takes almost 30\% of the original MBT when applying CFS-BA for feature selection. Due to the huge amount of data with high dimensionality, the ensemble method with the original features takes approximately 500s and 1000s for the AWID and CIC-IDS2017 datasets separately. Thanks to the feature selection method, the ensemble model with CFS-BA has mitigated the MBT considerably compared with that using all original features, all the MBTs of CFS-BA-Ensemble model on these three datasets have been restricted within 100s. Especially for the CIC-IDS2017 dataset, there is a significant reduction on the MBT of the ensemble classifier when using the CFS-BA based feature selection method, from 977.94s to 98.42s.

\subsubsection{Comparison with other feature selection methods}

\begin{figure*}
\centering

  \subfigure[]{\includegraphics[width=2.8in]{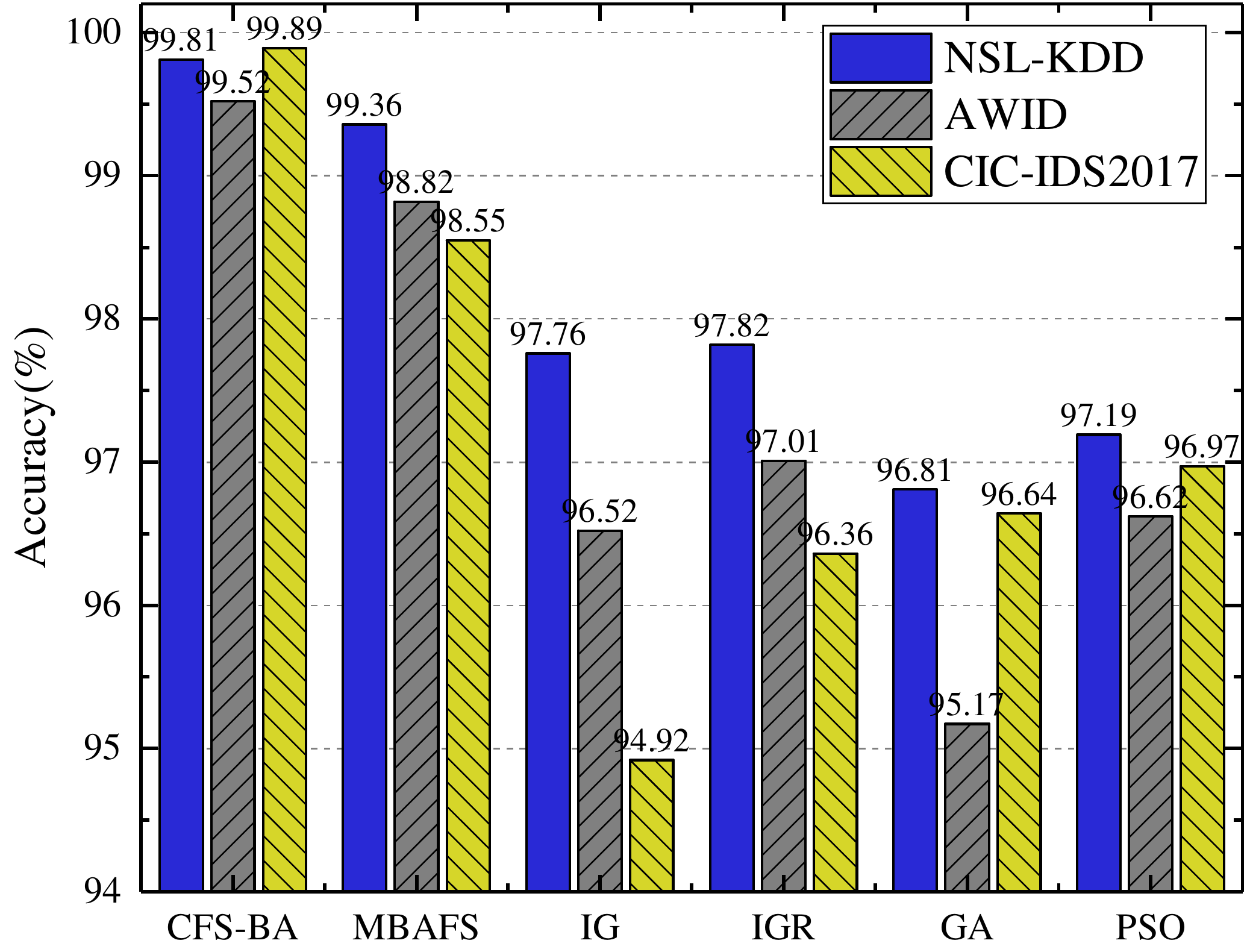}
  \label{fig3(a)}}
  \quad
  \hspace{.5in}
  \subfigure[]{\includegraphics[width=2.8in]{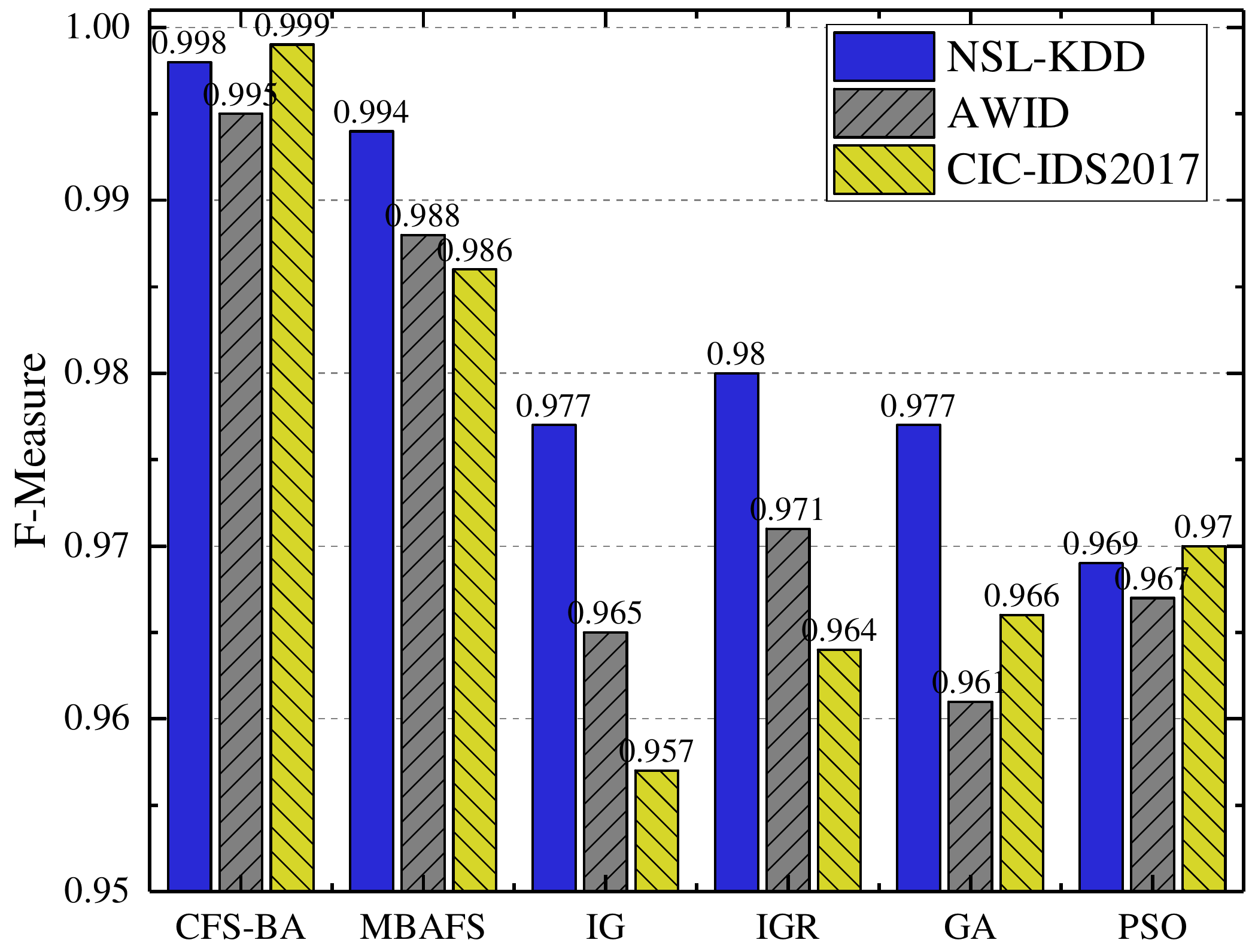}
  \label{fig3(b)}}

  \subfigure[]{\includegraphics[width=2.8in]{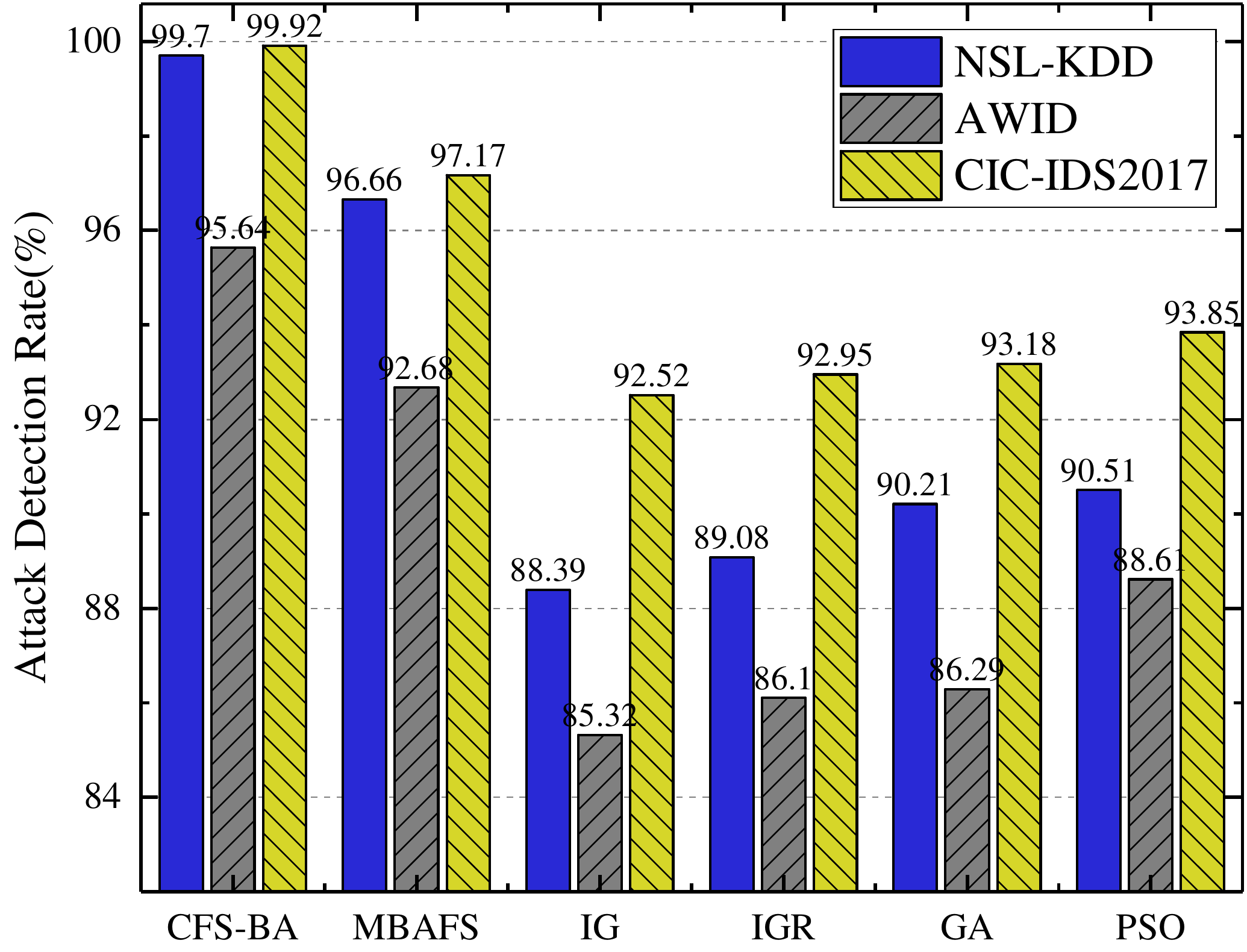}
  \label{fig3(c)}}
  \quad
  \hspace{.62in}
  \subfigure[]{\includegraphics[width=2.68in]{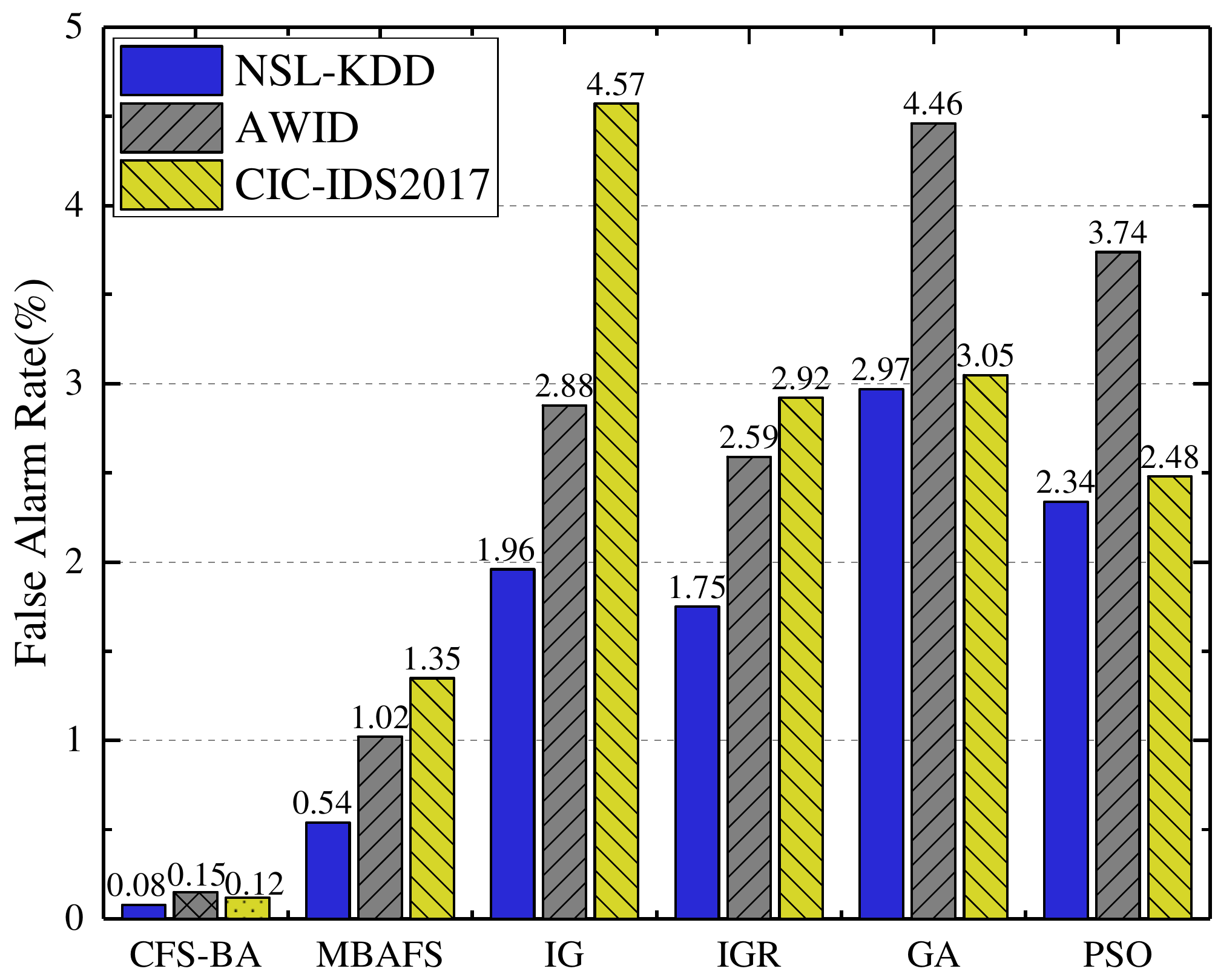}
  \label{fig3(d)}}

  \hspace{.03in}
  \subfigure[]{\includegraphics[width=2.74in]{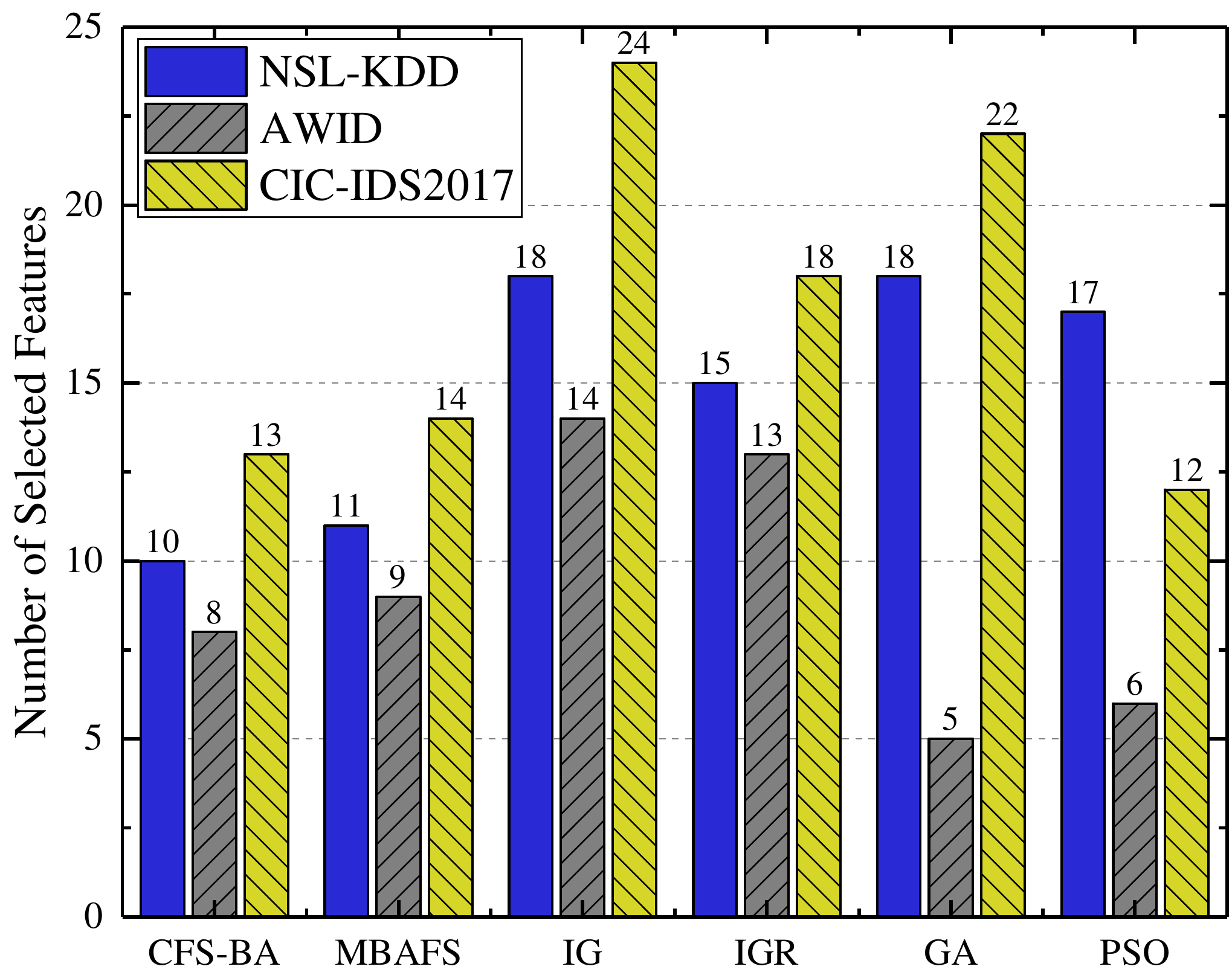}
  \label{fig3(e)}}
  \quad
  \hspace{.58in}
  \subfigure[]{\includegraphics[width=2.74in]{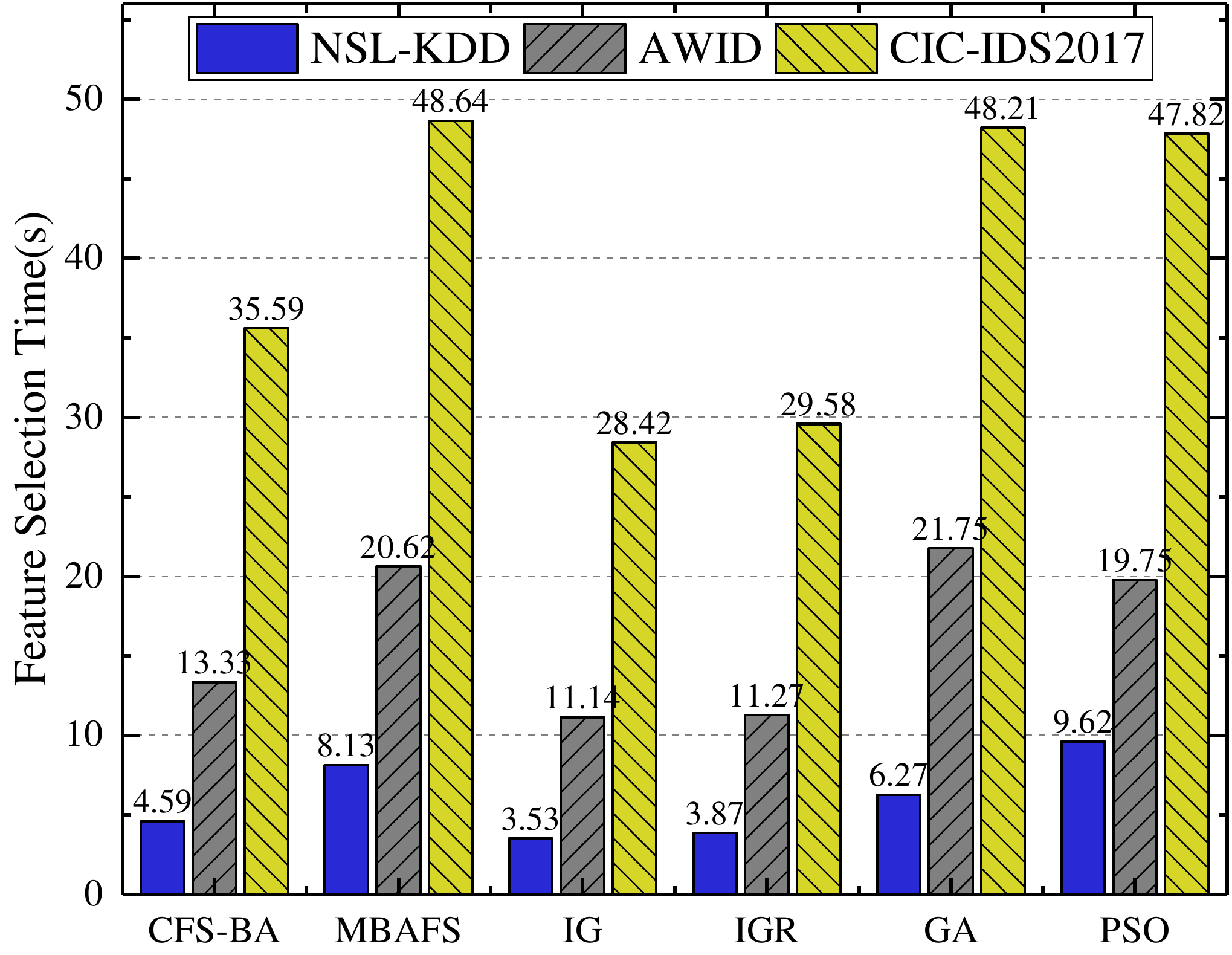}
  \label{fig3(f)}}

\caption{Comparison performance with other feature selection methods based on the three datasets.}
\label{fig3}
\end{figure*}

As explained in Section~\ref{subsection4.1}, the benchmark datasets reflect a contemporary and complex threat environment. The increased number of attack classes and its highly imbalanced records pose a significant challenge to every machine learning approach. In order to further evaluate our proposed IDS model, we compare it with some well-known feature selection methods, namely IG (Information Gain)~\cite{azhagusundari2013feature}, IGR (Information Gain Ratio)~\cite{mansouri2019intrusion}, GA (Genetic Algorithm)~\cite{pal2017genetic}, PSO (Particle Swarm Optimization)~\cite{zhang2015comprehensive}, and MBAFS (Modified Bat Algorithm for Feature Selection)~\cite{yang2017feature} by conducting experiments based on these three datasets. Likewise, in this comparative study we use the common metrics in the context of Acc, F-Measure, ADR, and FAR. Especially, to figure out the efficiency of the proposed IDS, the comparison has also been done in terms of number of selected features and its selection time. Fig.~\ref{fig3} summarizes the average performance of our model as compared to the other feature selection methods based on the same proposed voting based ensemble classifier.

First, as shown in Fig.~\ref{fig3(a)}, the accuracy of our proposed model outperforms that of other algorithms based feature selection in every dataset, and the proposed CFS-BA-Ensemble approach achieves the highest average accuracy rate of 99.81\%, 99.52\%, and 99.89\% over the NSL-KDD, AWID, and CIC-IDS2017 datasets, respectively. Similarly, Fig.~\ref{fig3(b)} indicates that our proposed model exhibits better F-Measure than other feature selection methods on all datasets through extracting more relevant feature subsets, which increase the value of F-Measure from 0.969 to 0.998, 0.961 to 0.995, and 0.957 to 0.999 over these three datasets. Next, the attack detection rate, which stands for the accuracy rate for the attack classes, is an important indicator to evaluate the performance of an IDS. According to Fig.~\ref{fig3(c)}, it can be observed that the attack detection rate of our proposed model ranges from 95.64\% to 99.92\%, which significantly exceeds other feature selection methods based on any one of the five sets. Moreover, as Fig.~\ref{fig3(d)} illustrates, our proposed CFS-BA based model achieves the lowest FAR values of 0.08\%, 0.15\%, and 0.12\% based on the NSL-KDD, AWID, and CIC-IDS2017 datasets separately. In comparison with other feature selection methods, our proposed model has mitigated FAR considerably on each dataset and guaranteed the effectiveness of an IDS.

Notably, Fig.~\ref{fig3(e)} and Fig.~\ref{fig3(f)} exhibit the number of selected features using different algorithms and its selection time, which can indicate the efficiency of an IDS. When compared to IG and IGR, although the proposed method takes a little more time than them, CFS-BA selects less features, and as seen in Fig.~\ref{fig3(a)}, the accuracy of the proposed IDS is much higher than that of IG and IGR. For GA and PSO based feature selection methods, each of them obtains less features than CFS-BA on the AWID dataset, however, they need more feature selection time on all the five sets and could not achieve better detection accuracy. MBAFS, a modified bat algorithm for feature selection, is considered to be most similar to our feature selection method. According to Fig.~\ref{fig3(a)}$-$\ref{fig3(d)}, MBAFS performs better than any of other methods in terms of these performance metrics except the proposed CFS-BA, and its performance is only slightly worse than our method. Since MBAFS introduces random bats and mutation mechanism, the search space is expanded in every iteration and the subset may be generated in any uncertain direction. Therefore, as Fig.~\ref{fig3(e)} illustrates, the subset selected by MBAFS contains one more feature than CFS-BA on any of these five sets, which may affect the performance of the IDS if the additional feature are not highly correlated. In addition, as the number of iterations before convergence increases, the feature selection time taken by MBAFS is more than ours, which can be seen in Fig.~\ref{fig3(f)}. In general, CFS-BA is superior to other feature selection methods in terms of performance and efficiency.

\subsubsection{Comparison with other classifiers}

Similarly, to evaluate the performance of our proposed ensemble classifier, experiments have been conducted using different classification algorithms among five preprocessed sets with CFS-BA feature selection. First, the proposed voting based ensemble classifier with AOP combination rule is chosen, and we construct a stacking classifier with C4.5, RF, and Forest PA as base classifiers, and Logistic Regression (LR)~\cite{Kleinbaum} as meta classifier to make a comparison with our voting classifier. Second, we select some widely studied ensemble algorithms, such as AdaBoost (AB)~\cite{hu2008adaboost} and Gradient Boosted Machine (GBM)~\cite{friedman2001greedy} to make a comparison. Third, some single classifiers like k-Nearest Neighbor (kNN)~\cite{liao2002use}, Classification and Regression Trees (CART)~\cite{breiman2017classification}, and Multi-Layer Perceptron (MLP)~\cite{leung1991complex} have been chosen as well.

\begin{figure*}
\centering
  \subfigure[]{\includegraphics[width=2.75in]{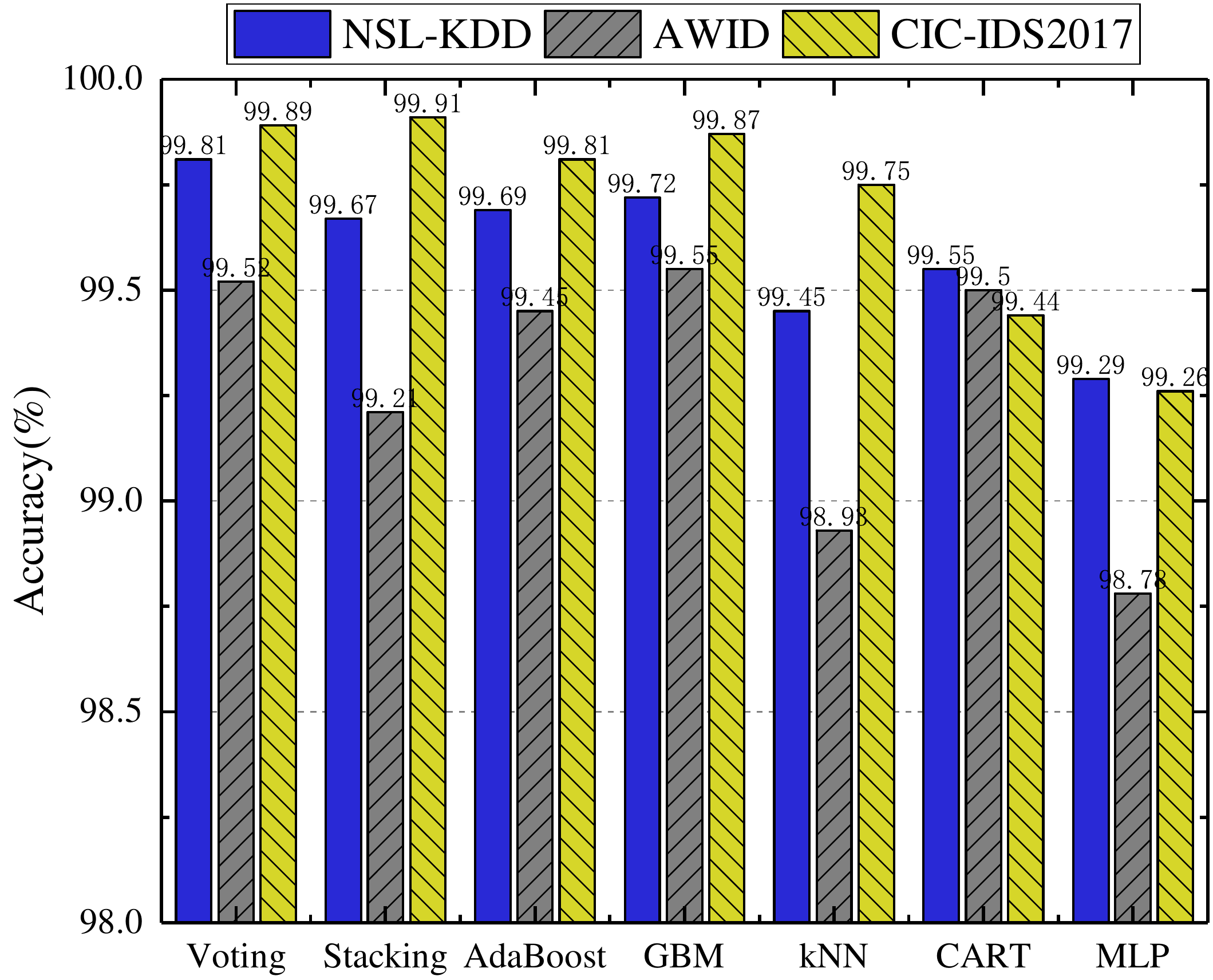}
  \label{fig6(a)}}
  \quad
  \hspace{.5in}
  \subfigure[]{\includegraphics[width=2.75in]{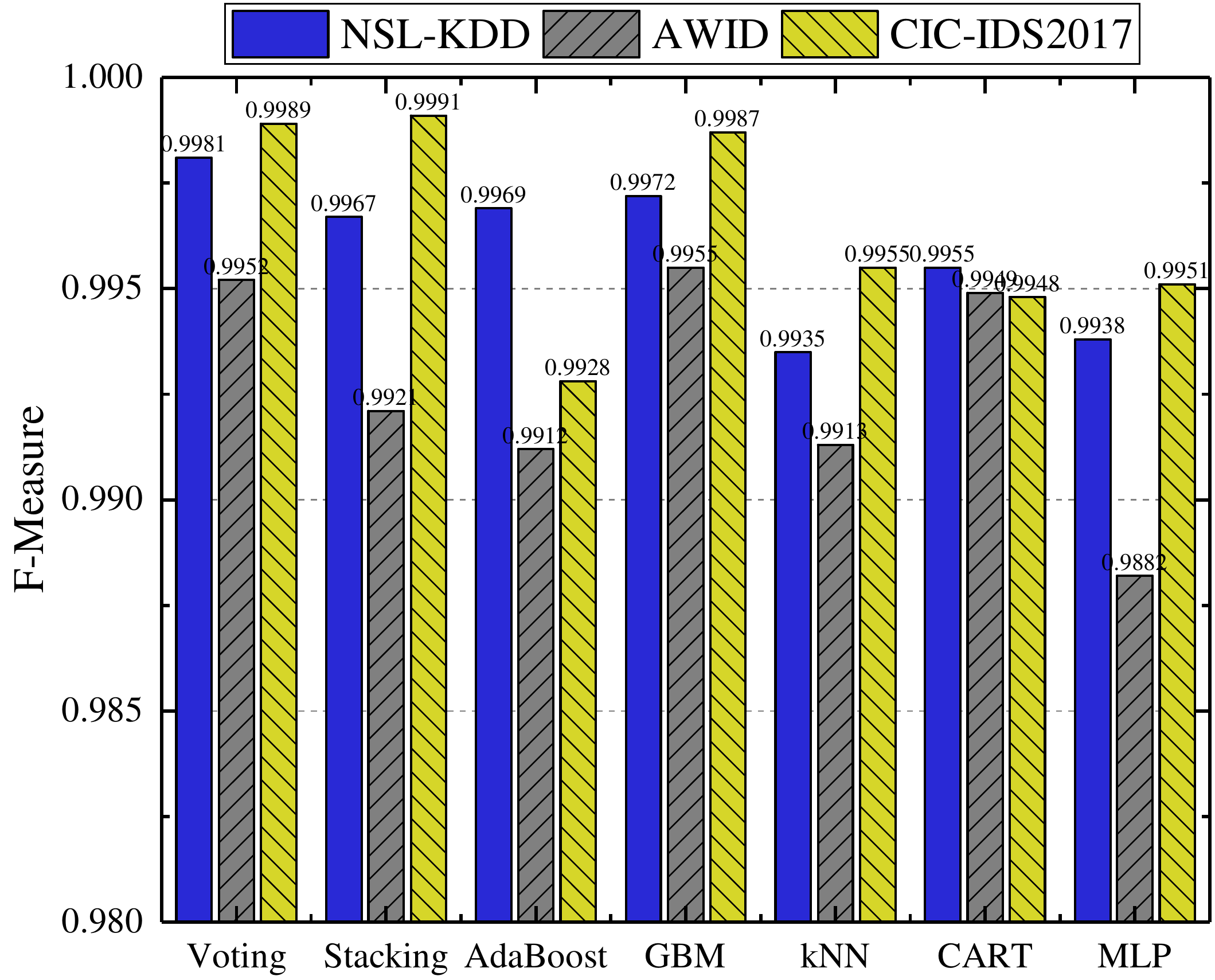}
  \label{fig6(b)}}

  \subfigure[]{\includegraphics[width=2.75in]{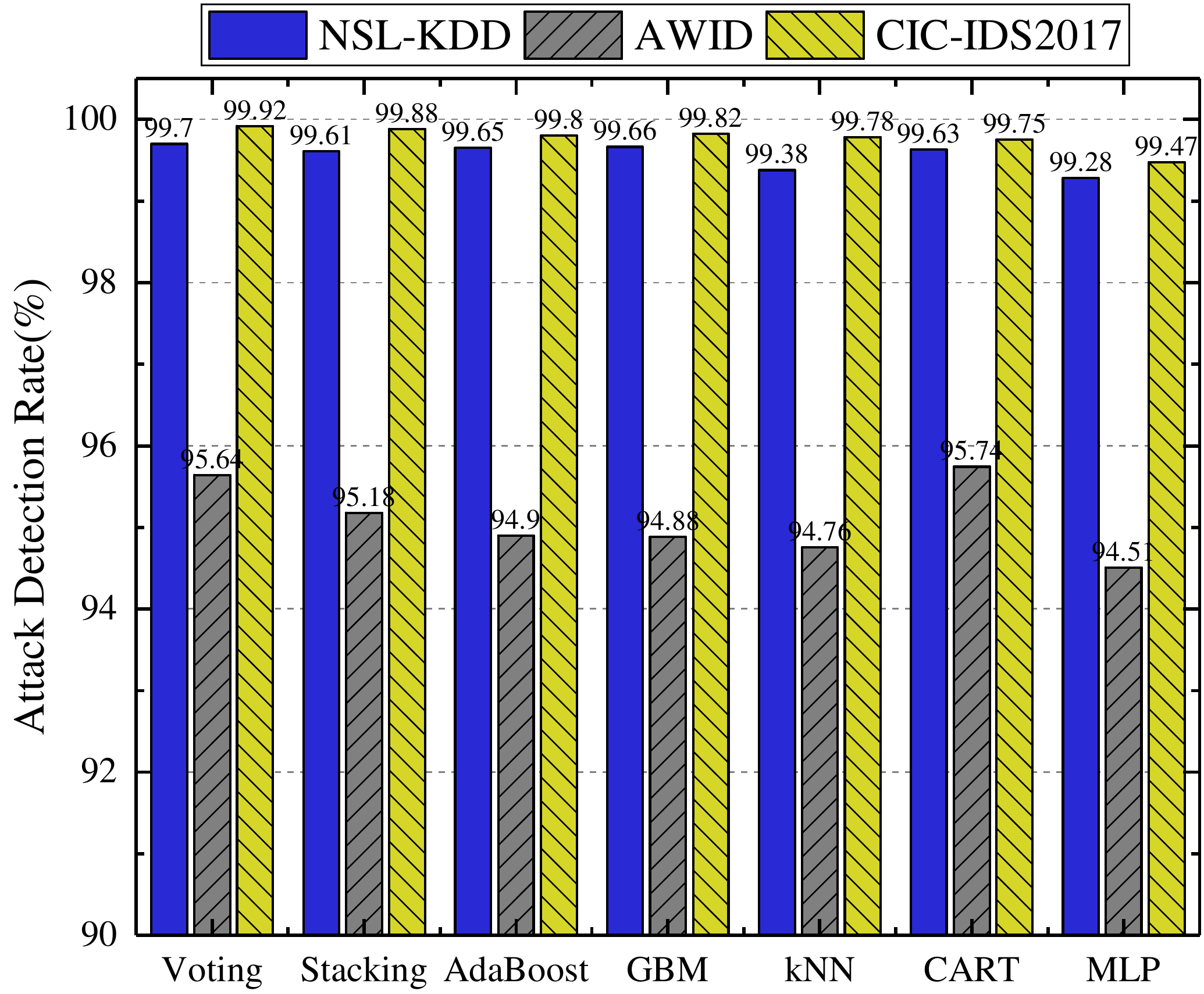}
  \label{fig6(c)}}
  \quad
  \hspace{.55in}
  \subfigure[]{\includegraphics[width=2.70in]{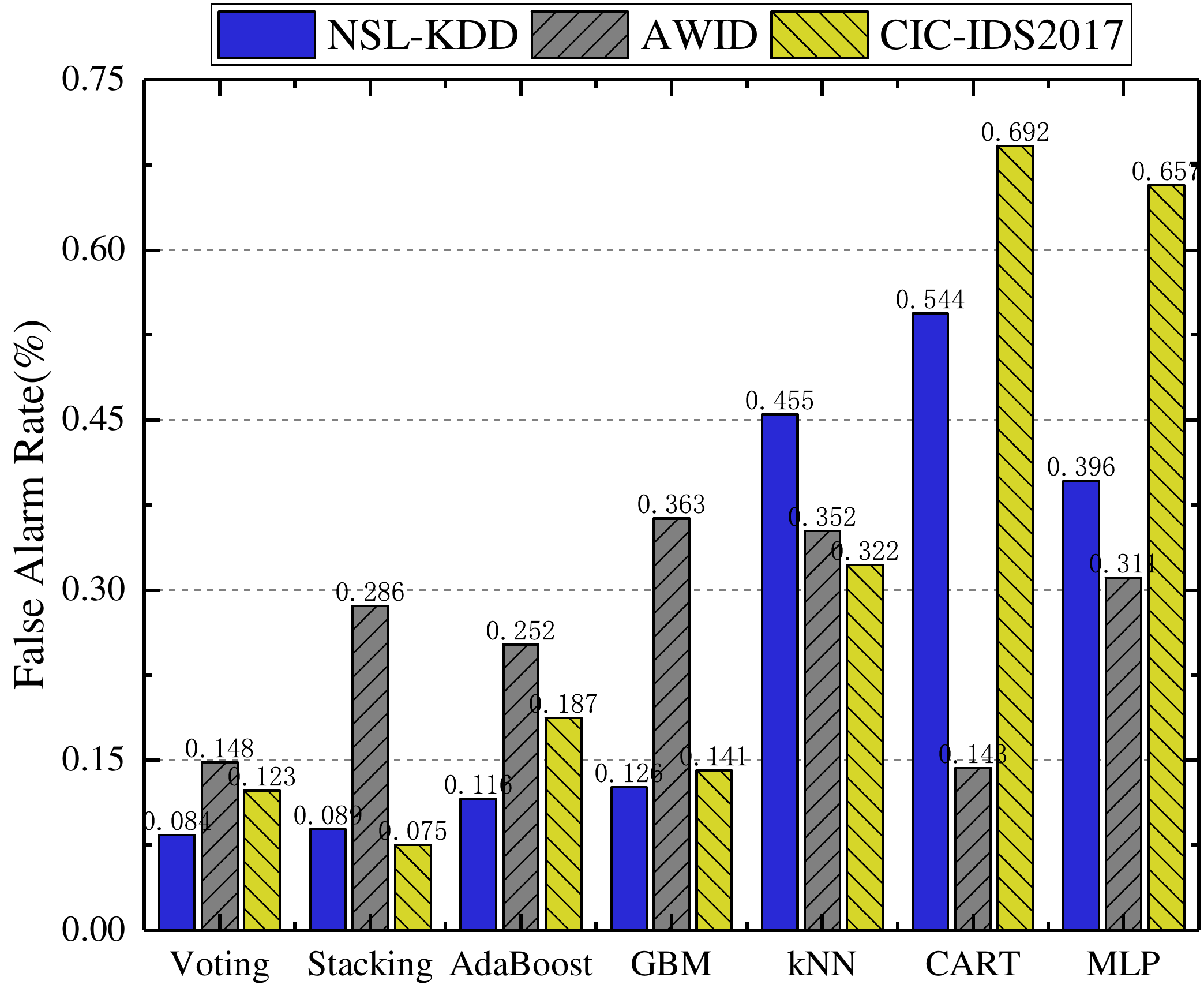}
  \label{fig6(d)}}

\caption{Comparison performance of per classifier across three datasets with 10f cross-validation.}
\label{fig6}
\end{figure*}

However, an algorithm may not be able to achieve good results over all datasets, which makes quite difficult to
compare different algorithms over multiple datasets. In order to perform the comparison of classifiers in a correct way~\cite{verma2019machine}, it is necessary to statistically analyze the significance of the classifiers' performance. Thus, the significance tests has been considered essential to find whether the classifiers are significantly different from each other or not~\cite{demvsar2006statistical}. In order to give a thoroughly comparative study, two statistical significance tests, Friedman test~\cite{friedman1937use} and Nemenyi post-hoc test~\cite{nemenyi1962distribution}, are adopted. In our case, the null-hypothesis is that there is no performance difference among different classifiers, and it can be rejected if at least one classifier is found significantly different from at least one other classifier. Since there are 7 classifiers to be compared in this case, Friedman test is chosen to prove whether at least one classifier performs significantly better than another one over all datasets~\cite{tama2019depth}. If the Friedman test reports a significant difference, to detect between which classifiers those differences appear, the Nemenyi post-hoc test will be then proceeded for pairwise multiple comparisons.

\begin{figure*}
\centering
  \subfigure[$\alpha$ = 0.05]{\includegraphics[width=2.8in]{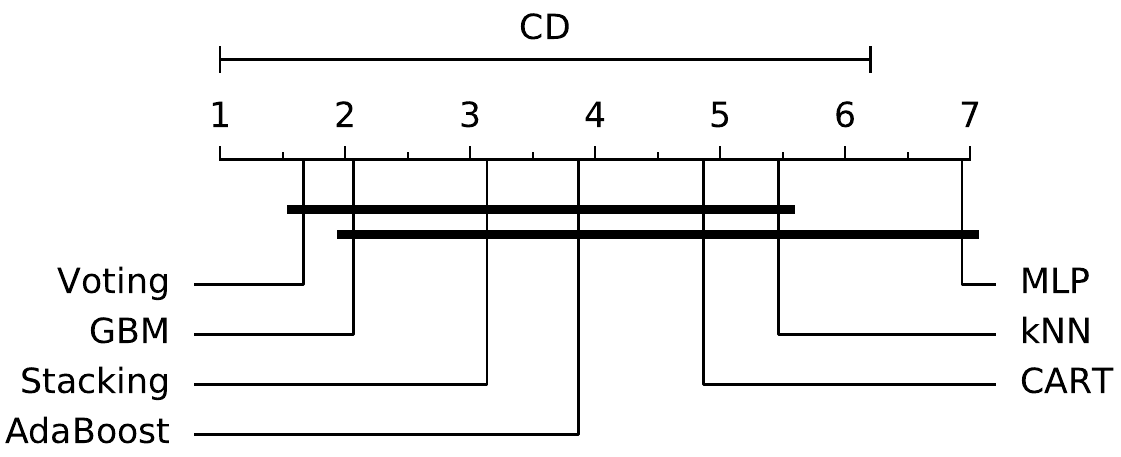}
  \label{fig7(a)}}
  \quad
  \hspace{.5in}
  \subfigure[$\alpha$ = 0.1]{\includegraphics[width=2.8in]{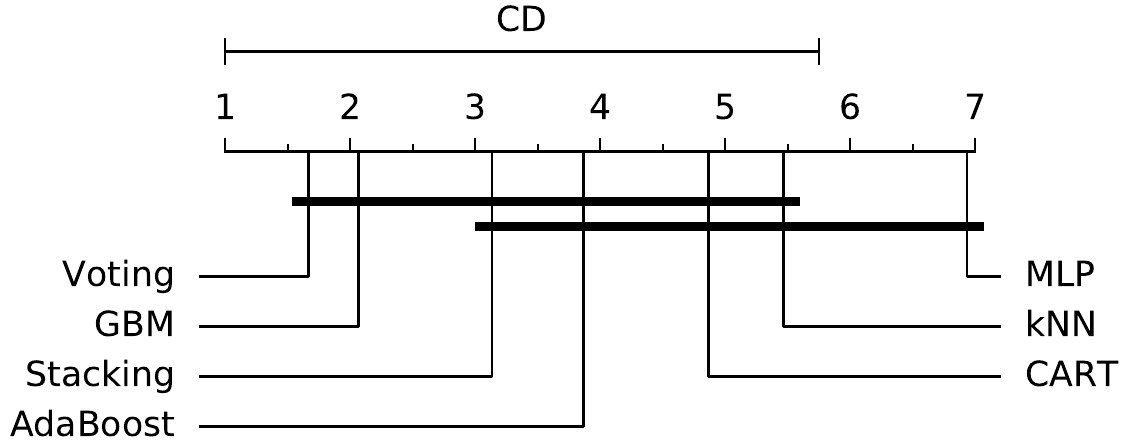}
  \label{fig7(b)}}
\caption{Critical difference of all classifiers in term of accuracy metric.}
\label{fig7}
\end{figure*}

\begin{figure*}
\centering
  \subfigure[$\alpha$ = 0.05]{\includegraphics[width=2.8in]{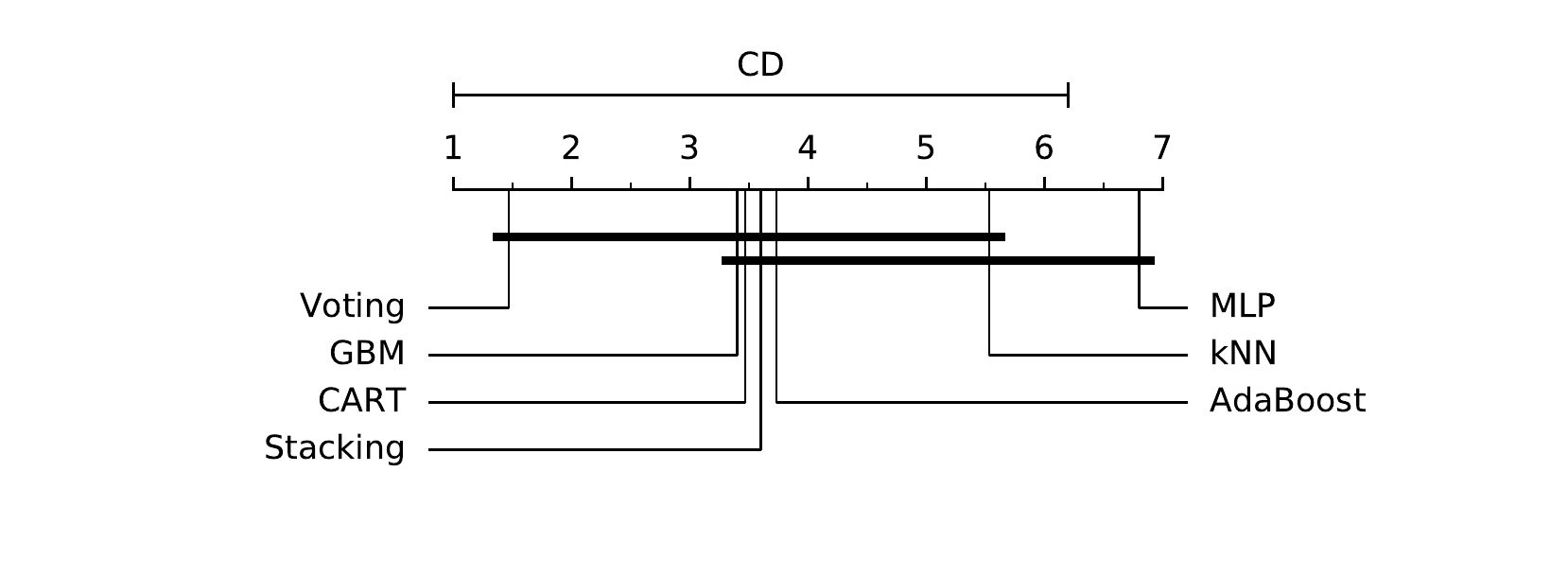}
  \label{fig8(a)}}
  \quad
  \hspace{.5in}
  \subfigure[$\alpha$ = 0.1]{\includegraphics[width=2.8in]{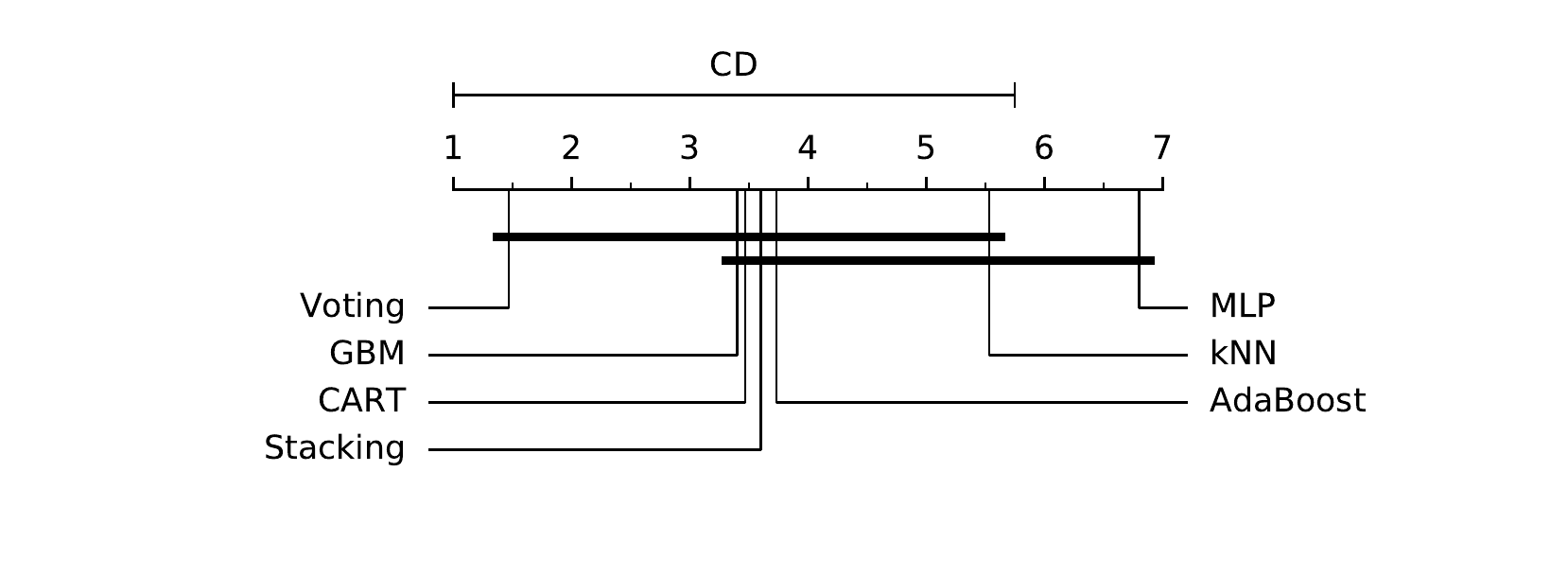}
  \label{fig8(b)}}
\caption{Critical difference of all classifiers in term of attack detection rate metric.}
\label{fig8}
\end{figure*}

\begin{table}[pos=t]
\centering
\caption{Average ranks for 10f cross-validation across three datasets.}
\label{table-mean_rank}
\begin{tabular}{@{}lllll@{}}
\toprule
         & Accuracy & F-Measure & ADR   & FAR   \\ \midrule
Voting   & 1.667    & 1.967     & 1.467 & 1.867 \\
Stacking & 3.133    & 2.933     & 3.600 & 2.733 \\
AdaBoost & 3.867    & 5.033     & 3.733 & 3.333 \\
GBM      & 2.067    & 2.367     & 3.400 & 4.000 \\
kNN      & 5.467    & 5.233     & 5.533 & 5.533 \\
CART     & 4.867    & 4.633     & 3.467 & 4.533 \\
MLP      & 6.933    & 5.867     & 6.800 & 6.000 \\\bottomrule
\end{tabular}
\end{table}

For the Friedman test, it ranks the algorithms for each dataset separately. For example, for a given dataset, the algorithm performing best gets the rank of 1, the second best gets rank 2, and so on. After that, Friedman test will do it again over another dataset until we obtain all rankings on all datasets. Let $r_{ij}$ be the rank of the $j$-th algorithm on the $i$-th dataset, where $i=1,2,...,n$, and $j=1,2,...,k$. Then, the average rank of $j$-th algorithm can be calculated as Eq.~\ref{equation16}, and the Friedman statistic (F-Statistic) can be computed as Eq.~\ref{equation18}, where $\chi^2_F$ is calculated as Eq.~\ref{equation17}.
\begin{gather}
	R_{j}= \frac{1}{n}\sum\limits_{i=1}^n r_{ij} \label{equation16}\\
	\chi^2_F=\frac{12n}{k(k+1)} \left[\sum\limits_{j=1}^k R_j^2 - \frac{k(k+1)^2}{4} \right]\label{equation17}\\
    F{-}Stastic=\frac{(n-1)\chi^2_F}{n(k-1)-\chi^2_F}\label{equation18}\\
    CD =q_{\alpha}\sqrt{k(k+1)\over 6n} \label{equation19}
\end{gather}

If the null-hypothesis is accepted, F-Statistic will be distributed according to the F-distribution for a given $\alpha$ with $k{-}1$ and $(k{-}1)(n{-}1)$ degrees of freedom. In this study, the values of $k$ and $n$ are set as 7 and 3, and two values of $\alpha$ (0.05 and 0.1) are considered. Otherwise, if we reject the null-hypothesis, then the Nemenyi post-hoc test will be performed to compare all classifiers with each other. The performance of two classifiers is significantly different when the difference between their average ranks is greater or equal to the critical difference (CD), where CD can be calculated as Eq.~\ref{equation19}. In Eq.~\ref{equation19}, $k$ and $n$ still represent the numbers of classifiers and datasets, and $q_{\alpha}$ is the critical value. Due to 7 classifiers are used for comparison, according to Table 5 (a) in~\cite{demvsar2006statistical}, $q_{0.05}=2.949$ and $q_{0.1}=2.693$ in our case.

\begin{table}[pos=t]
\centering
\caption{Friedman test statistics for 10f cross-validation.}
\label{table-F-test}
\begin{tabular}{@{}lllll@{}}
\toprule
            & Accuracy & F-Measure & ADR    & FAR    \\ \midrule
F-Statistic & 6.5665   & 2.0810    & 3.3242 & 1.7904 \\
\textit{p} value     & 0.0029   & 0.1319    & 0.0363 & 0.1839 \\
$\alpha = 0.05$      & R        & A         & R      & A      \\
$\alpha = 0.1$       & R        & A         & R      & A      \\ \bottomrule
\end{tabular}
\end{table}

First, we analyze the average value of all mentioned metrics achieved with 10f cross-validation on the three datasets, which is shown in Fig.~\ref{fig6}. It is observed from Fig.~\ref{fig6(a)} that Voting, GBM, and Stacking outperform other classifiers in terms of accuracy (99.81\%, 99.55\%, and 99.91\%) over NSL-KDD, AWID, and CIC-IDS2017 datasets separately but MLP achieves lowest accuracy values across all datasets. Similarly, Voting, GBM, and Stacking perform best in terms of F-Measure according to Fig.~\ref{fig6(b)}. However, kNN, MLP, and AdaBoost perform worst with F-Measure of 0.9935, 0.9882, and 0.9928. In terms of ADR metric, our proposed Voting based ensemble classifier performs best by achieving 99.7\% and 99.92\% on the NSL-KDD and CIC-IDS2017 datasets, and CART achieves the highest ADR value (95.74\%) on the AWID dataset, whereas MLP is the worst performer over the three datasets. Fig.~\ref{fig6(d)} indicates the average FAR values of all classifiers over all three datasets. Voting, CART, and Stacking separately exhibit the lowest FAR values of 0.084, 0.143, and 0.075 for NSL-KDD, AWID, and CIC-IDS2017 datasets. However, CART achieves the worst performance in terms of FAR for the NSL-KDD and CIC-IDS2017 datasets, and the worst performing classifier is GBM for the AWID dataset.

Then, the performance results are statistically assessed using Friedman and Nemenyi post-hoc test. According to experimental results, the average ranks of all the classifiers for 10f cross-validation are shown in Table~\ref{table-mean_rank}. Thus, the F-Statistic and $p$ value for each performance metric is computed, and Table~\ref{table-F-test} shows Friedman test statistics for 10f cross-validation results. From the results it is observed that $p$ values under accuracy and ADR are less than 0.05, therefore the null-hypothesis is rejected and we can conclude that the performance of the classifiers is significantly different in terms of accuracy and ADR metrics. In order to detect which classifier pairs perform significantly different, Nemenyi post-hoc test is performed, and the results of the pairwise comparison over accuracy and ADR values are presented in Fig.~\ref{fig7} and Fig.~\ref{fig8}. It is indicated that for accuracy metric the classifier's performance is highly significant (shown in Fig.~\ref{fig7(a)}) in the case of Voting-MLP and less significant (shown in Fig.~\ref{fig7(b)}) in the case of GBM-MLP, whereas remaining pairs are not significant. As shown in Fig.~\ref{fig8}, the classifier's ADR measure is only found highly significant in case of Voting-MLP pair, while all other pairs are not significant. The experimental results show that Voting and GBM are suitable classifiers if the IDS demands high accuracy, and we highly suggest our Voting based ensemble classifier due to it also shows outstanding performance in terms of ADR metric.

\subsubsection{Comparison with other combination rules}
\begin{table*}
\centering
\caption{Comparison of different combination rules under the NSL-KDD dataset based on accuracy.}
\label{table-rule-nsl}
\begin{tabular}{@{}llllll@{}}
\toprule
       & Average of probabilities & Majority voting & Product of probabilities & Minimum probability & Maximum probability \\ \midrule
Normal & \textbf{99.90}                    & 99.75           & 98.67                    & 98.67               & 98.71               \\
DoS    & \textbf{99.92}                    & 99.63           & 99.48                    & 99.48               & 99.32               \\
PRB    & \textbf{99.48}                    & 98.53           & 97.32                    & 97.32               & 98.69               \\
R2L    & \textbf{94.57}                    & 93.38           & 90.55                    & 90.15               & 89.61               \\
U2R    & 55.77                    & \textbf{57.69}           & 53.85                    & 53.85               & 51.92               \\ \bottomrule
\end{tabular}
\end{table*}

\begin{table*}[pos=!t]
\centering
\caption{Comparison of different combination rules under the AWID dataset based on accuracy.}
\label{table-rule-awid}
\begin{tabular}{@{}llllll@{}}
\toprule
              & Average of probabilities & Majority voting & Product of probabilities & Minimum probability & Maximum probability \\ \midrule
Normal        & \textbf{99.85}           & 99.75           & 98.67                    & 98.67               & 98.71               \\
Injection     & \textbf{99.98}           & 99.90           & 98.91                    & 98.91               & 99.15               \\
Flooding      & \textbf{92.71}           & 90.16           & 88.92                    & 89.29               & 86.45               \\
Impersonation & 93.21                    & 91.89           & 89.45                    & 89.45               & \textbf{93.63}      \\ \bottomrule
\end{tabular}
\end{table*}

In this section, we explain the experimental results using CFS-BA-Ensemble approach with different combination rules we reached during the experiments. Similarly, the average accuracy values of outputs from 10 iterations of 10f validation approach are used for evaluation of the models. As mentioned in Section~\ref{subsection_vote}, minimum probability, maximum probability, majority voting, product of probabilities, and average of probabilities are common combination rules when using voting technique to construct an ensemble classifier. Therefore, in order to evaluate the multi-classification performance of these aggregation methods, from Table~\ref{table-rule-nsl} to Table~\ref{table-rule-cic}, we compare and analyze the average accuracy values of each combination rule for each attack type of different datasets.

Table~\ref{table-rule-nsl} shows the accuracy values of each rule for the NSL-KDD dataset. For `Normal', `DoS', `PRB', and `R2L', the average of probabilities combination rule achieves the highest performance accuracy values of 99.90\%, 99.92\%, 99.48\%, and 94.57\% compared to the other combination rules. Although the performance of majority voting rule is better than average of probabilities in `U2R' attack, the improvement in accuracy may lead to only one more attack instance identified due to the number of `U2R' instances in the NSL-KDD dataset. Therefore, compared to majority voting, we prefer to use average of probabilities combination rule for more accurate classification on most instances of the NSL-KDD dataset.

According to the results shown in Table~\ref{table-rule-awid}, the highest accuracy values of 99.85\%, 99.98\%, and 92.71\% are obtained for `Normal', `Injection', and `Flooding' with average of probabilities combination rule based on the AWID dataset. For the `Impersonation' attack, the performance of maximum probability rule is slightly better than AOP but, when we consider most of the cases and the difference between accuracy values for two cases, we still suggest the average of probabilities rule for the AWID dataset. Similarly, as shown in Table~\ref{table-rule-cic}, it is obviously observed that the best performance is still achieved with the average of probabilities combination rule for most classes of the CIC-IDS2017 dataset, such as `Benign', `DoS slowloris', `DoS Hulk', and `DoS GoldenEye'. The majority voting combination rule achieves the highest accuracy of 99.02\% for `DoS Slowhttptest', however, it produces the worst accuracy of 97.77\% for `DoS Hulk' and has no advantages for other types of attacks when compared to the average of probabilities combination rule. According to results in this table, it can also be indicated that the performance of the maximum probability combination rule for `Heartbleed' (90.91\%) is better than the other rules. However, as seen in Table~\ref{table2}, there are only 11 instances of `Heartbleed' contained in the CIC-IDS2017 dataset. The advantage on the classification for `Heartbleed' attacks cannot make up for its drawbacks on the other attacks when in comparison with the average of probabilities combination rule.

Based on the experimental results on the three datasets, therefore, in this study, we decided to apply the AOP as combination rule in our proposed CFS-BA-Ensemble model.

\begin{table*}[pos=!t]
\centering
\caption{Comparison of different combination rules under CIC-IDS2017 dataset based on accuracy.}
\label{table-rule-cic}
\begin{tabular}{@{}llllll@{}}
\toprule
                 & Average of probabilities & Majority voting & Product of probabilities & Minimum probability & Maximum probability \\ \midrule
Benign           & \textbf{99.88}           & 99.73           & 97.80                    & 97.82               & 97.34               \\
DoS slowloris    & \textbf{99.26}           & 98.34           & 97.48                    & 97.48               & 96.64               \\
DoS Slowhttptest & 98.95                    & \textbf{99.02}  & 97.29                    & 97.29               & 96.02               \\
Dos Hulk         & \textbf{99.97}           & 97.77           & 98.56                    & 97.80               & 97.77               \\
DoS GoldenEye    & \textbf{99.59}           & 99.10           & 97.64                    & 97.64               & 97.15               \\
Heartbleed       & 81.82                    & 81.82           & 72.73                    & 72.73               & \textbf{90.91}      \\ \bottomrule
\end{tabular}
\end{table*}

\subsubsection{Comparison with the state of the art methods}

\begin{figure*}
\centering
  \subfigure[KDDTest+]{\includegraphics[width=2.5in]{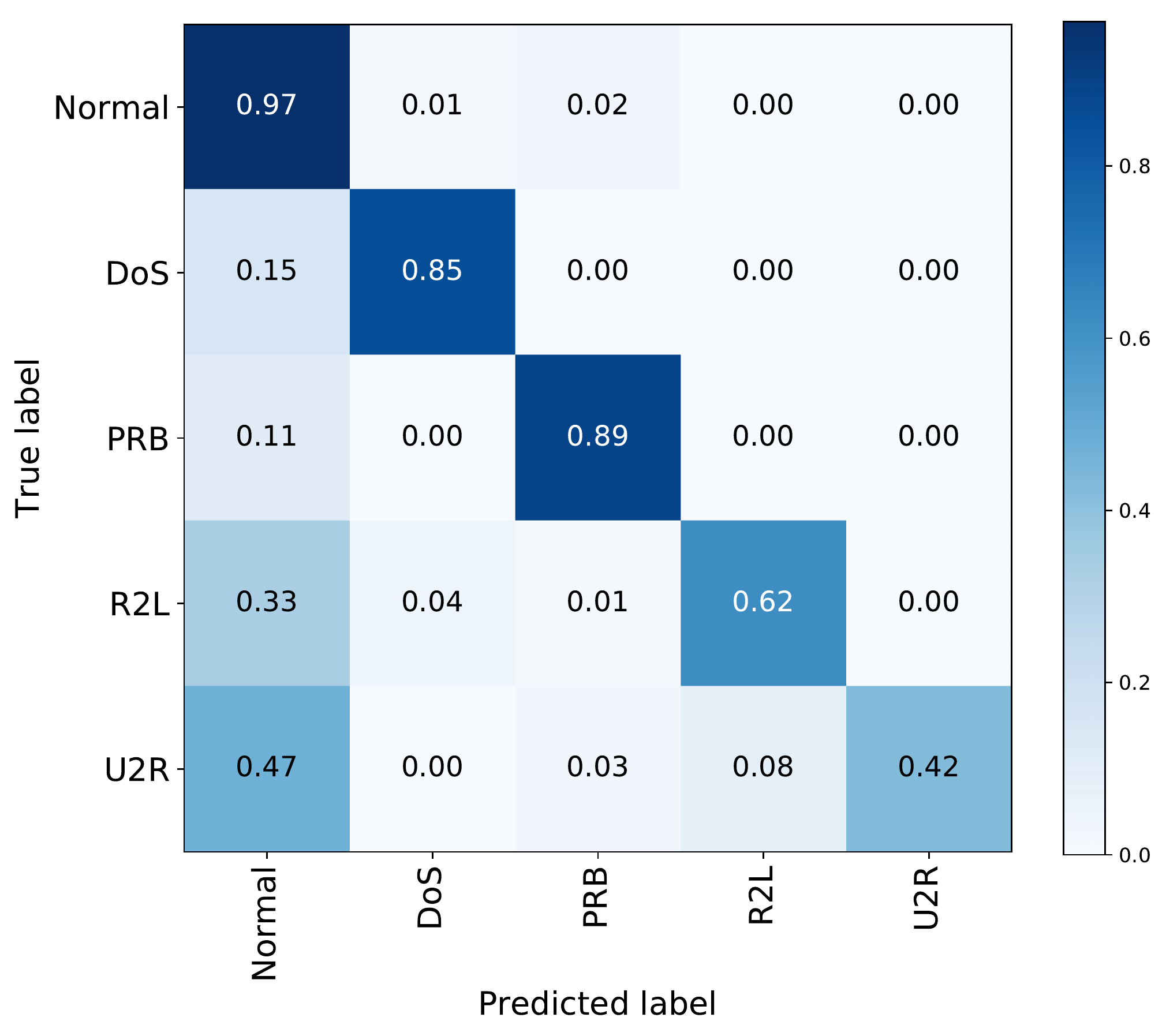}
  \label{fig_matrix2(a)}}
  \quad
  \hspace{.8in}
  \subfigure[KDDTest-21]{\includegraphics[width=2.5in]{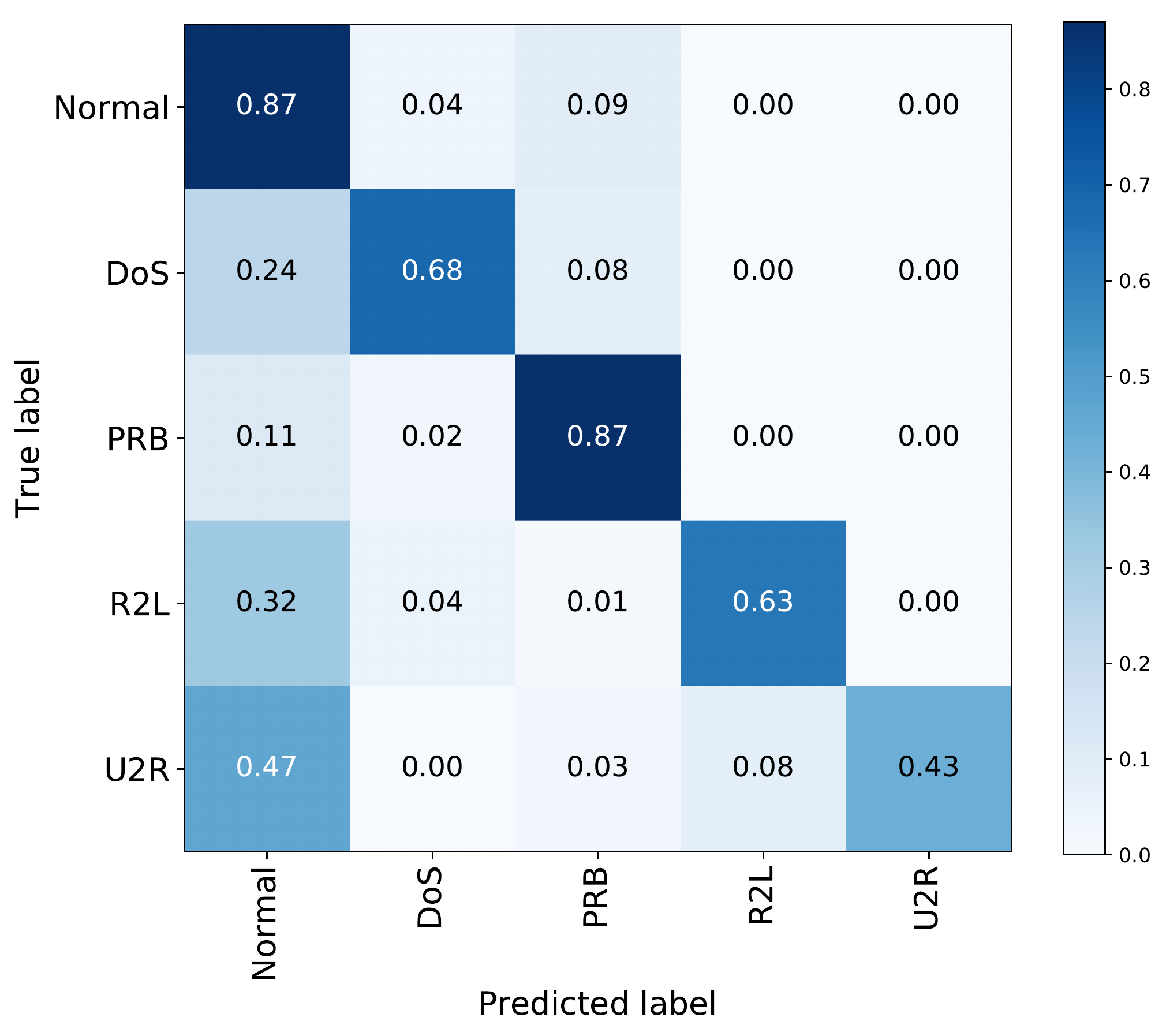}
  \label{fig_matrix2(b)}}

\caption{Normalized confusion matrices based on the KDDTest+ and KDDTest-21 sets.}
\label{fig_matrix2}
\end{figure*}

NSL-KDD dataset provides some different test sets, such as KDDTest+ and KDDTest-21 for benchmarking the machine learning algorithms. To evaluate the proposed model on unseen attacks, we have conducted experiments with the proposed CFS-BA-Ensemble model by using the datasets KDDTrain+ for training and KDDTest+ and KDDTest-21 for testing. Fig.~\ref{fig_matrix2} indicates the multi-class classification performance of the proposed IDS based on the KDDTest+ and KDDTest-21 test sets. As seen in Fig.~\ref{fig_matrix2(a)}, the proposed model can achieve the accuracy of 0.97 for normal traffic, whereas 0.85 and 0.89 for `DoS' and `PRB' attacks based on the KDDTest+ set. Similarly, it can be observed from Fig.~\ref{fig_matrix2(b)} that our method can obtain the accuracy of 0.87, 0.68, and 0.87 for `Normal', `DoS', and `PRB' instances. Moreover, our proposed method obtains accuracy values of 0.62 and 0.42 for the `R2L' and `U2R' attacks on KDDTest+, which are similar to that on KDDTest-21. On one hand, the proposed model has been trained on the KDDTrain+ set, where the `R2L' and `U2R' instances occupy the smallest proportion of all instances. On the other hand, the `R2L' and `U2R' instances are the same in the KDDTest+ and KDDTest-21 set according to Table~\ref{table1}, so the classification performance for them is almost the same.

To extend the benchmark, we also have compared our CFS-BA-Ensemble with the performance achieved by previous studies that use the datasets KDDTest+ and KDDTest-21 for testing. The comparison results with some of the existing approaches on these two sets are shown in Table~\ref{table-kdd-test}. The highest detection accuracy is achieved by the proposed approach based on the experimental results on KDDTest+, which outperforms the other recent IDS techniques, including FSSL~\cite{ashfaq2017fuzziness}, FSSL-EL~\cite{gao2018novel}, and TSE-IDS~\cite{tama2019tse}. Besides having superior detection accuracy, the proposed method also outperforms significantly other approaches in terms of detection rate metric. Even though EM-FS~\cite{pham2018improving} performs best in terms of FAR metric, it only achieves the accuracy of 84.25\% based on 35 features. However, our proposed method can obtain higher accuracy of 87.37\% with FAR of 3.19\% based on only 10 features, which still outperforms EM-FS to some extent. Moreover, according to the experimental results tested on the KDDTest-21 set, the proposed approach can achieve the accuracy of 73.57\%, DR of 73.6\%, and FAR of 12.92\% with a 10-feature subset, which clearly outperforms other state of the art classifiers in terms of all the evaluation metrics shown in Table~\ref{table-kdd-test}.

\begin{table*}[]
\centering
\caption{Comparison results with other existing methods on KDDTest+ and KDDTest-21.}
\label{table-kdd-test}
\begin{threeparttable}
\begin{tabular}{@{}llllllll@{}}
\toprule
Method      & Dataset           & Feature selection & Classification method    & \# Features & Acc(\%) & DR(\%)  & FAR(\%) \\ \midrule
NBTree~\cite{tavallaee2009detailed} & KDDTest+ & N/A  & NB Tree               & 41          & 82.02   & N/A     & N/A     \\
Fuzzy~\cite{kromer2011fuzzy}         & KDDTest+ & N/A  & Fuzzy classifiers     & 41          & 82.74   & 86.7    & 3.9     \\
SVM~\cite{pervez2014feature}         & KDDTest+ & N/A  & SVM                   & 41          & 82.37   & 82      & 15     \\
FS+GAR-forest~\cite{kanakarajan2016improving}& KDDTest+ & Symmetrical & GAR-forest & 32      & 85.056   & 85.1    & 12.2     \\
TDTC~\cite{pajouh2016two}            & KDDTest+ & LDA+PCA & NB+CF-kNN              & N/A     & 84.86   & N/A     & 4.86     \\
FSSL~\cite{ashfaq2017fuzziness}      & KDDTest+ & Clustering & FSSL                & 41      & 84.12   & N/A      & N/A     \\
EM-FS~\cite{pham2018improving}       & KDDTest+ & IGR & Bagging(C4.5)              & 35      & 84.25   & N/A     & \textbf{2.79}\\
FSSL-EL~\cite{gao2018novel}          & KDDTest+ & PCA & Ensemble(CART)             & 20      & 84.54   & N/A      & 5.31     \\
TSE-IDS~\cite{tama2019tse}           & KDDTest+ & Hybrid & Two-stage Ensemble      & 37      & 85.797  & 86.8     & 11.7     \\
NBTree~\cite{tavallaee2009detailed} & KDDTest-21& N/A  & NB Tree                   & 41          & 66.16   & N/A     & N/A     \\
FSSL~\cite{ashfaq2017fuzziness}      & KDDTest-21& Clustering & FSSL                & 41          & 68.82   & N/A      & N/A     \\
FSSL-EL~\cite{gao2018novel}          & KDDTest-21& PCA & Ensemble(CART)             & 20          & 71.29   & N/A     & 20.35     \\
TSE-IDS~\cite{tama2019tse}           & KDDTest-21& Hybrid & Two-stage Ensemble      & 37          & 72.52  & 72.5     & 18.00     \\
Proposed    & KDDTest+         & CFS-BA            & Voting(C4.5,RF,ForestPA) & \textbf{10} & \textbf{87.37} & \textbf{87.4} & 3.19 \\
Proposed    & KDDTest-21       & CFS-BA            & Voting(C4.5,RF,ForestPA) & \textbf{10} & \textbf{73.57} & \textbf{73.6}& \textbf{12.92} \\ \bottomrule
\end{tabular}
\begin{tablenotes}
\item[] N/A: name not available.
\end{tablenotes}
\end{threeparttable}
\end{table*}

In order to further interpret the advantages of the proposed approach, in this section, some state of the art studies applied on each dataset are compared with our proposed CFS-BA-Ensemble method. More precisely, the comparison includes the feature selection techniques, the classification method, the number of selected features, accuracy, FAR, and DR for intrusion detection. Furthermore, to compare more fairly with these existing methods, we ensure that the datasets used by these methods, even if the specific version of the datasets, are the same as ours. Similarly, these existing methods all adopt cross-validation approach. The results of our proposed method compared to the existing approaches in each dataset are presented in Tables~\ref{table-binary} and \ref{table-multiple}.

Although the multi-class classification performance of our proposed method has been proven through experiments, to provide more reference for the readers, we still compare the results of our CFS-BA-Ensemble method with other earlier researches in binary classification based on NSL-KDD, AWID, and CIC-IDS2017 datasets, which is shown in Table~\ref{table-binary}. First of all, it can be seen in Table~\ref{table-binary} that our proposed model outperforms other similar ensemble classifiers, such as FS-EL~\cite{tama2017extensive}, XGBoost-IDS~\cite{bansal2018extreme}, and TSE-IDS~\cite{tama2019tse} when using 10f cross-validation as a validation technique. There are also some deep learning methods for IDS in the current literature such as DEMISe~\cite{parker2019demise}, DeepWindow~\cite{shi2019deepwindow}, and HELAD~\cite{zhong2020helad}. Even though HELAD performs very well in terms of accuracy rate, DR, and FAR, the proposed method can still achieve a better performance compared with these methods. When compared to these binary classification methods, the proposed CFS-BA-Ensemble method has a slight advantage on accuracy and DR against all of them applied on the three datasets. And although some of previous studies did not provide data for FAR, it achieves extremely low FAR by 0.08\%, 0.15\%, and 0.12\% across all datasets, which is a useful property for real-world IDSs. Additionally, the proposed method may not be the best by considering the number of selected features, however, it is worth selecting only a few more features to effectively improve the performance of the classifier.

As shown in the following Table~\ref{table-multiple}, we then compare the performance of our proposed method, CFS-BA-Ensemble, with these existing methods for multi-class attack classification. For the NSL-KDD dataset, OR+FS~\cite{sainis2018feature} exhibits a high accuracy value of 99.43\% based on the selected 6 features, however, the proposed approach achieves higher accuracy by 0.38\% with 4 more features needed. Similarly, in contrast to earlier ensemble methods on AWID and CIC-IDS2017 datasets, like MVWIDS~\cite{alotaibi2016majority}, ELWNIDS~\cite{vaca2018ensemble}, and DARE~\cite{attak2018application}, the proposed approach achieves better performances in accuracy and DR while limiting FAR at a lower level.

\begin{table*}[]
\centering
\caption{Comparison results with other state of the art binary classification approaches.}
\label{table-binary}
\begin{threeparttable}
\begin{tabular}{@{}llllllll@{}}
\toprule
Method      & Dataset           & Feature selection & Classification method    & \# Features & Acc(\%) & DR(\%)  & FAR(\%) \\ \midrule
FRCM~\cite{panigrah2016fuzzy}        & KDDTrain+         & Greedy Stepwise   & Fuzzy Ownership NN       & 11          & 99.6356      & 99.6145 & 0.309   \\
FS-EL~\cite{tama2017extensive} & KDDTrain+         & CFS+PSO           & Boosting(CART)           & 11          & 99.7285      & 99.77   & N/A     \\
OneR-BN~\cite{panigrahi2019481}     & KDDTrain+         & OneR              & BN+TAN                   & N/A         & 99.7412      & 99.7646 & 0.2792  \\
TSE-IDS~\cite{tama2019tse}           & KDDTrain+        & Hybrid            & Two-stage Ensemble       & 37          & 96.388       & N/A    & N/A    \\
DEMISe~\cite{parker2019demise}        & AWID-CLS-R        & Autoencoder,MI & RBFC           & \textbf{7}          & 98        & 99.04   & 3    \\
SSLA~\cite{ran2019semi}        & AWID-CLS-R        & N/A               & Ladder Network           & 95          & 99.28        & 99.45   & 0.23    \\
DARE~\cite{attak2018application}     & CIC-IDS2017(Wed.) & N/A               & One-class SVM              & \textbf{10}          & 66        & 57   & N/A    \\
XGBoost-IDS~\cite{bansal2018extreme}     & CIC-IDS2017(Wed.) & N/A               & XGBoost                       & 80          & 91.36           & 98.38         & 12     \\
ZED-IDS~\cite{catillo2019discovery}     & CIC-IDS2017(Wed.) & N/A               & Autoencoder              & 83          & 95.73        & 95.82   & 4.32    \\
DeepWindow~\cite{shi2019deepwindow}  & CIC-IDS2017(Wed.) & MI+MIC            & LSTM                     & N/A         & 99.5         & 99.4    & N/A     \\
HELAD~\cite{zhong2020helad}          & CIC-IDS2017(Wed.) & DIS+DBN           & Autoencoder+LSTM          & 50        & 99.58         & 99.58    & 2.15     \\
Proposed    & KDDTrain+         & CFS-BA            & Voting(C4.5,RF,ForestPA) & \textbf{10}          & \textbf{99.81}        & \textbf{99.8}    & \textbf{0.08}     \\
Proposed    & AWID-CLS-R        & CFS-BA            & Voting(C4.5,RF,ForestPA) & 8           & \textbf{99.52}        & \textbf{99.5}    & \textbf{0.15}    \\
Proposed    & CIC-IDS2017(Wed.) & CFS-BA            & Voting(C4.5,RF,ForestPA) & 13          & \textbf{99.89}        & \textbf{99.9}    & \textbf{0.12}    \\ \bottomrule
\end{tabular}
\begin{tablenotes}
\item[] N/A: name not available.
\end{tablenotes}
\end{threeparttable}
\end{table*}

\begin{table*}[]
\centering
\caption{Comparison results with other state of the art multi-class classification approaches.}
\label{table-multiple}
\begin{threeparttable}
\begin{tabular}{@{}llllllll@{}}
\toprule
Method   & Dataset           & Feature selection & Classification method    & \# Features & Acc(\%) & DR(\%)     & FAR(\%) \\ \midrule
AR-C4.5~\cite{chae2014feature}  & KDDTrain+         & Attribute Ratio   & C4.5                     & 22          & 99.794       & N/A        & N/A     \\
SS-BN~\cite{zhang2013effective}    & KDDTrain+         & Sequential Search & Bayesian Network         & 11          & 98.98        & N/A        &0.60         \\
OR+FS~\cite{sainis2018feature}    & KDDTrain+         & IQR,CFS+BFS       & kNN                      & \textbf{6}           & 99.43        & N/A        & N/A     \\
IG-RT~\cite{thanthrige2016machine}   & AWID-CLS-R        & IG               & Random Tree    & 41          & 95.12        & 92         & 0.538     \\
MVWIDS~\cite{alotaibi2016majority}   & AWID-CLS-R        & N/A               & Voting(ET,RF,Bagging)    & 20          & 96.32        & 96         & N/A     \\
ELWNIDS~\cite{vaca2018ensemble}  & AWID-CLS-R        & CFS               & RF                       & 18          & 99.096       & N/A        & 0.248   \\
DARE~\cite{attak2018application}     & CIC-IDS2017(Wed.) & N/A               & RF                       & \textbf{10}          & 98           & 98         & N/A     \\
DeepDetect~\cite{asad2019deepdetect}     & CIC-IDS2017(Wed.) & N/A               & ANN                       & 80          & 98.694           & 98.694         & 1.882     \\
XGBoost-IDS~\cite{bansal2018extreme}     & CIC-IDS2017(Wed.) & N/A               & XGBoost                       & 80          & 99.54           & 99.54         & 0.15     \\
Proposed & KDDTrain+         & CFS-BA            & Voting(C4.5,RF,ForestPA) & 10          & \textbf{99.81}        & \textbf{99.8}       & \textbf{0.08}     \\
Proposed & AWID-CLS-R        & CFS-BA            & Voting(C4.5,RF,ForestPA) & \textbf{8}           & \textbf{99.52}        & \textbf{99.5}       & \textbf{0.15}    \\
Proposed & CIC-IDS2017(Wed.) & CFS-BA            & Voting(C4.5,RF,ForestPA) & 13          & \textbf{99.89}        & \textbf{99.9}       & \textbf{0.12}    \\ \bottomrule
\end{tabular}
\begin{tablenotes}
\item[] N/A: name not available.
\end{tablenotes}
\end{threeparttable}
\end{table*}

In general, the proposed method achieves promising results in the context of accuracy rate, DR, and FAR across the NSL-KDD, AWID, and CIC-IDS2017 datasets in comparison with the earlier studies. However, it should be noted that Table~\ref{table-binary} and \ref{table-multiple} just provide a snapshot of comparison between our proposed model and the state of the art methods in intrusion detection problem. Thus, there might be some limitations in this comparison. For example, data processing method, parameter setting of the algorithm, and many other experimental factors are all unknown for the existing techniques. Although we have tried to make as fair a comparison as possible, we cannot claim that our proposed intrusion detection model always performs better when compared to any of the other methods in the context of intrusion detection. However, according to the comparison results indicated in Table~\ref{table-binary} and \ref{table-multiple}, our proposed CFS-BA-Ensemble method still provides a powerful competitive advantage in the intrusion detection domain.

\section{Conclusions}\label{section5}
Although many machine learning approaches have been proposed to increase the efficacy of IDSs, it is still a problem for existing intrusion detection algorithms to achieve good performance. In this paper, to deal with the high-dimensional and unbalanced network traffic, we propose a novel intrusion detection framework, which is based on the feature selection and ensemble learning techniques. First, we propose a CFS-BA algorithm with the aim of selecting the optimal subset based on the correlation between features. Then, the ensemble classifier based on C4.5, RF, and ForestPA with the AOP rule is introduced to construct the classification model. Finally, the proposed IDS is evaluated by 10f cross-validation over three intrusion detection datasets.

The experimental results are promising with an accuracy of classification equal to 99.81\%, 99.8\% DR and 0.08\% FAR with a subset of 10 features for the NSL-KDD dataset, and the obtained results for the AWID provide accuracy of 99.52\% and 0.15\% FAR with a subset composed of only 8 features. Remarkably, our model achieves the highest accuracy of 99.89\% and DR of 99.9\% on the subset of 13 features for the CIC-IDS2017 dataset. Then, the comparison with no feature selection method demonstrates encouraging performance on several metrics, and it should be noted that our proposal sharply reduces the MBT from 977.94s to 98.42s on the CIC-IDS2017 dataset. Our method also outperforms related feature selection approaches in terms of Acc, F-Measure, ADR, and efficiency while limiting FAR at relatively low levels. In addition, our solution shows outstanding performance in terms of ADR metric when compared to other classification algorithms, and the comparison results with the state of the art methods indicate that the proposed CFS-BA-Ensemble method can provide a powerful competitive advantage in the intrusion detection domain. Although the proposed CFS-BA Ensemble method has indicated superior performance, in the future work, its capability could be further improved to deal with rare attacks from the massive network traffic.

\section*{Acknowledgment}

This work is supported by National Key Research and Development Program of China under Grant No. 2018YFB\\1800602 and No. 2017YFB0801703, CERNET Innovation Project (NGIICS20190101, NGII20170406), and Ministry of Education-China Mobile Research Fund Project (MCM20\\180506).

\printcredits


\bibliographystyle{cas-model2-names}

\bibliography{cas-refs}

\bio{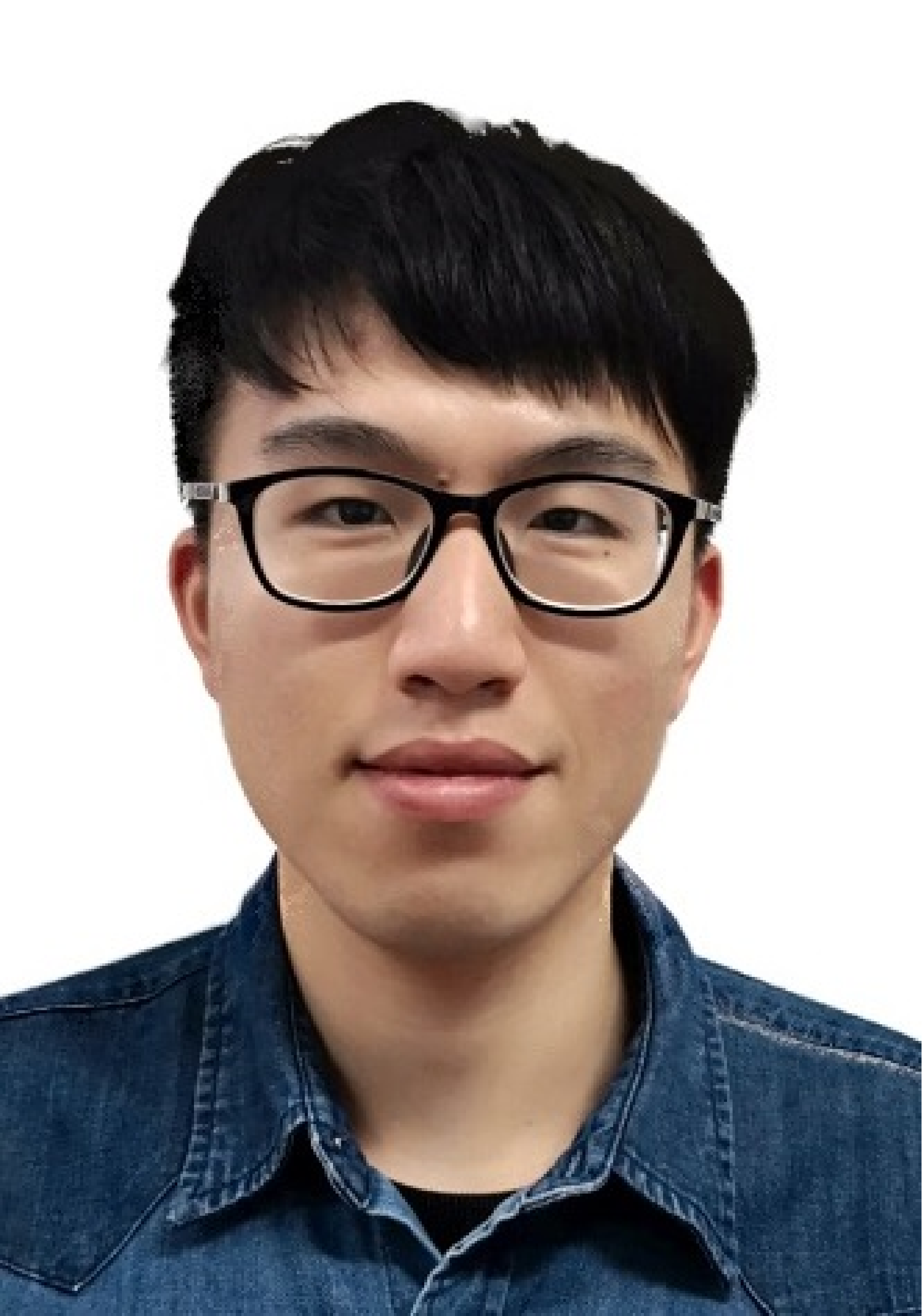}
Yuyang Zhou is currently pursuing the Ph.D. degree with the Cyber Science and Engineering School, Southeast University. His research interests include cyber security, traffic classification, and moving target defense.
\endbio
\vskip70pt
\bio{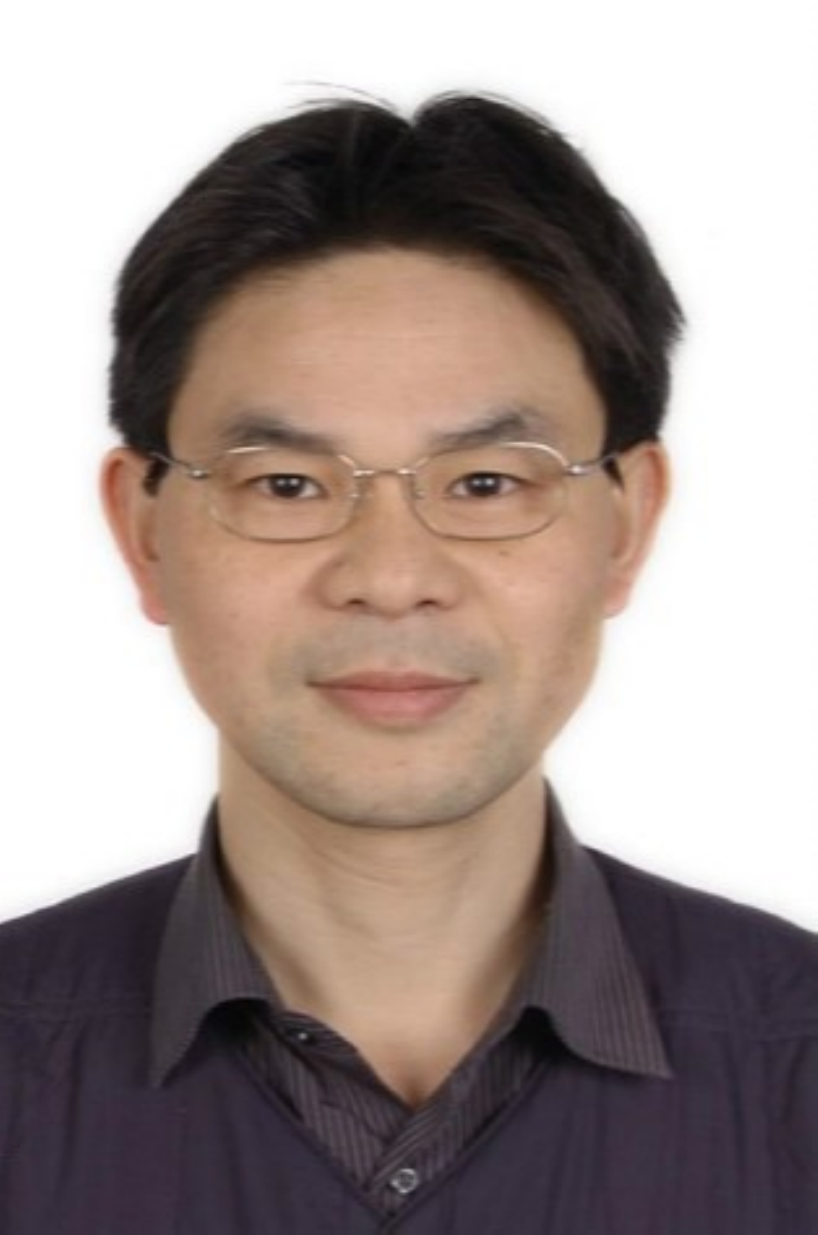}
Guang Cheng received the BS degree in Traffic Engineering from Southeast University in 1994, the MS degree in Computer Application from Heifei University of Technology in 2000, and the Ph.D degree in Computer Network from Southeast University in 2003. He is a full professor in the School of Cyber Science and Engineering, Southeast University, Nanjing, China. He is a senior member of the IEEE. His research interests include network security, network measurement and traffic behavior analysis.
\endbio
\vskip15pt
\bio{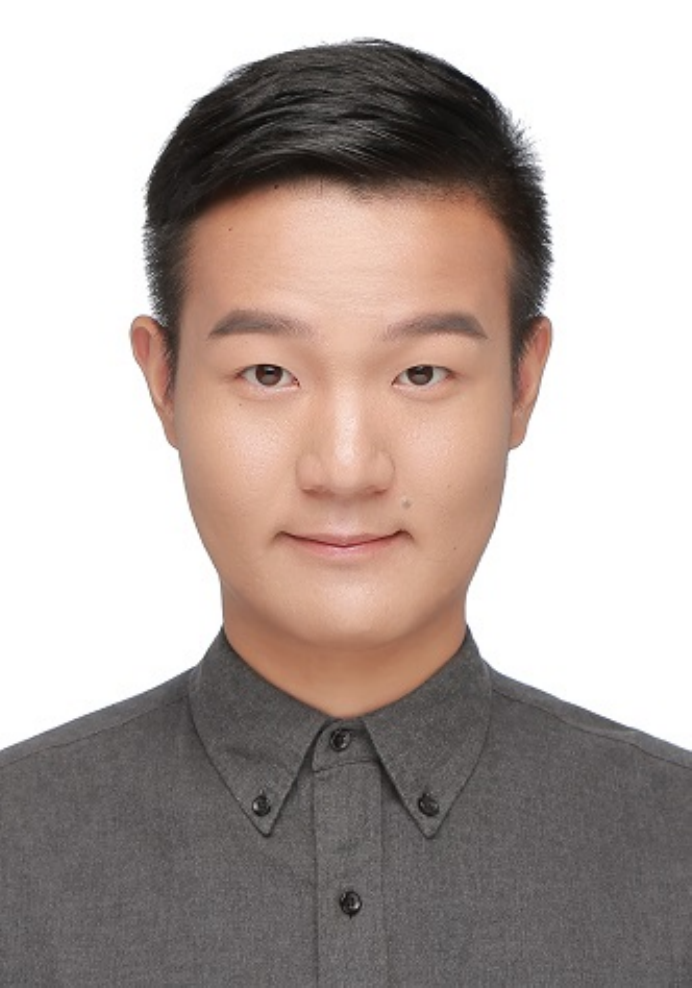}
Shanqing Jiang is currently pursuing the Ph.D. degree with the Cyber Science and Engineering School, Southeast University. His research interests include cyber security, traffic classification, and active defense.
\endbio
\vskip75pt
\bio{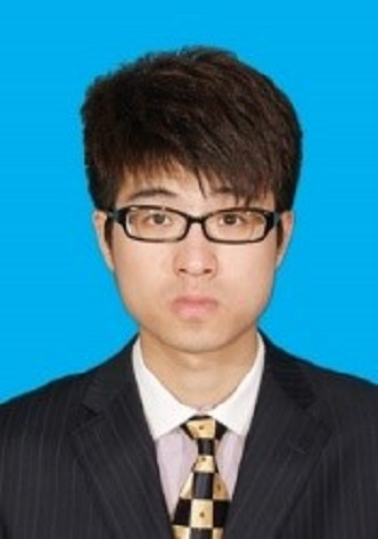}
Mian Dai is currently pursuing the Ph.D. degree with the Cyber Science and Engineering School, Southeast University. His research interests include traffic classification, network measurement and software defined network.
\endbio
\end{document}